\pdfoutput=1
\documentclass[pre,twocolumn,10pt,aps,longbibliography,superscriptaddress,nofootinbib]{revtex4-1}

\usepackage[dvipsnames]{xcolor}
\usepackage{amsmath,amssymb}
\usepackage{graphicx}
\usepackage{footmisc}
\usepackage{bbold,soul}
\usepackage{lipsum}
\usepackage{enumitem}
\usepackage{hyperref}
\usepackage{physics}

\makeatletter
\newcommand\footnoteref[1]{\protected@xdef\@thefnmark{\ref{#1}}\@footnotemark}
\makeatother

\newcommand{\vecr}{\mathbf{r}}
\newcommand{\vecx}{\mathbf{x}}

\newcommand{\vecy}{\mathbf{y}}
\newcommand{\vecz}{\mathbf{z}}

\newcommand{\vecw}{\mathbf{w}}
\newcommand{\Mprime}{{M'}}

\newcommand{\code}[1]{\texttt{#1}}
\usepackage{listings}

\definecolor{codegreen}{rgb}{0,0.6,0}
\definecolor{codegray}{rgb}{0.5,0.5,0.5}
\definecolor{codepurple}{rgb}{0.58,0,0.82}
\definecolor{orange}{rgb}{1, 0.13, 0.2}
\definecolor{backcolour}{rgb}{0.97,0.97,0.97}

\lstdefinestyle{mystyle}{
    backgroundcolor=\color{backcolour},   
    commentstyle=\color{codegreen},
    keywordstyle=\color{blue},
    numberstyle=\tiny\color{codegray},
    stringstyle=\color{orange},
    basicstyle=\footnotesize,
    breakatwhitespace=false,         
    breaklines=true,                 
    captionpos=b,                    
    keepspaces=true,                 
    numbers=left,                    
    numbersep=5pt,                  
    showspaces=false,                
    showstringspaces=false,
    showtabs=false,                  
    tabsize=2,
    basicstyle=\footnotesize\ttfamily
}

\definecolor{darkblue}{rgb}{0.0,0,0.75}
\hypersetup{
   colorlinks=true,
	linkcolor=red,
	filecolor=black,      
	urlcolor=darkblue,
	citecolor=red,
}

\begin{document}

\title{Uncertainty-aware predictions of molecular X-ray absorption spectra \\ using neural network ensembles}

\author{Animesh Ghose}
\affiliation{Computational Science Initiative, Brookhaven National Laboratory, Upton, New York 11973, USA}

\author{Mikhail Segal}
\affiliation{Computational Science Initiative, Brookhaven National Laboratory, Upton, New York 11973, USA}

\author{Fanchen Meng}
\affiliation{Center for Functional Nanomaterials, Brookhaven National Laboratory, Upton, New York 11973, USA}

\author{Zhu Liang}
\affiliation{Center for Functional Nanomaterials, Brookhaven National Laboratory, Upton, New York 11973, USA}

\author{Mark S. Hybertsen}
\affiliation{Center for Functional Nanomaterials, Brookhaven National Laboratory, Upton, New York 11973, USA}

\author{Xiaohui Qu}
\affiliation{Center for Functional Nanomaterials, Brookhaven National Laboratory, Upton, New York 11973, USA}

\author{Eli Stavitski}
\affiliation{National Synchrotron Light Source II, Brookhaven National Laboratory, Upton, New York 11973, USA}

\author{Shinjae Yoo}
\affiliation{Computational Science Initiative, Brookhaven National Laboratory, Upton, New York 11973, USA}

\author{Deyu Lu}
\email{dlu@bnl.gov}
\affiliation{Center for Functional Nanomaterials, Brookhaven National Laboratory, Upton, New York 11973, USA}

\author{Matthew R. Carbone}
\email{mcarbone@bnl.gov}
\affiliation{Computational Science Initiative, Brookhaven National Laboratory, Upton, New York 11973, USA}

\date{\today}

\begin{abstract}
As machine learning (ML) methods continue to be applied to a broad scope of problems in the physical sciences, uncertainty quantification is becoming correspondingly more important for their robust application. Uncertainty aware machine learning methods have been used in select applications, but largely for scalar properties. In this work, we showcase an exemplary study in which neural network ensembles are used to predict the X-ray absorption spectra of small molecules, as well as their point-wise uncertainty, from local atomic environments. The performance of the resulting surrogate clearly demonstrates quantitative correlation between errors relative to ground truth and the predicted uncertainty estimates. Significantly, the model provides an upper bound on the expected error. Specifically, an important quality of this uncertainty-aware model is that it can indicate when the model is predicting on out-of-sample data. This allows for its integration with large scale sampling of structures together with active learning or other techniques for structure refinement. Additionally, our models can be generalized to larger molecules than those used for training, and also successfully track uncertainty due to random distortions in test molecules. While we demonstrate this workflow on a specific example, ensemble learning is completely general. We believe it could have significant impact on ML-enabled forward modeling of a broad array of molecular and materials properties.
\end{abstract}

\maketitle

\section{Introduction}

\begin{figure*}[!htb]
    \centering
    \includegraphics[width=2\columnwidth]{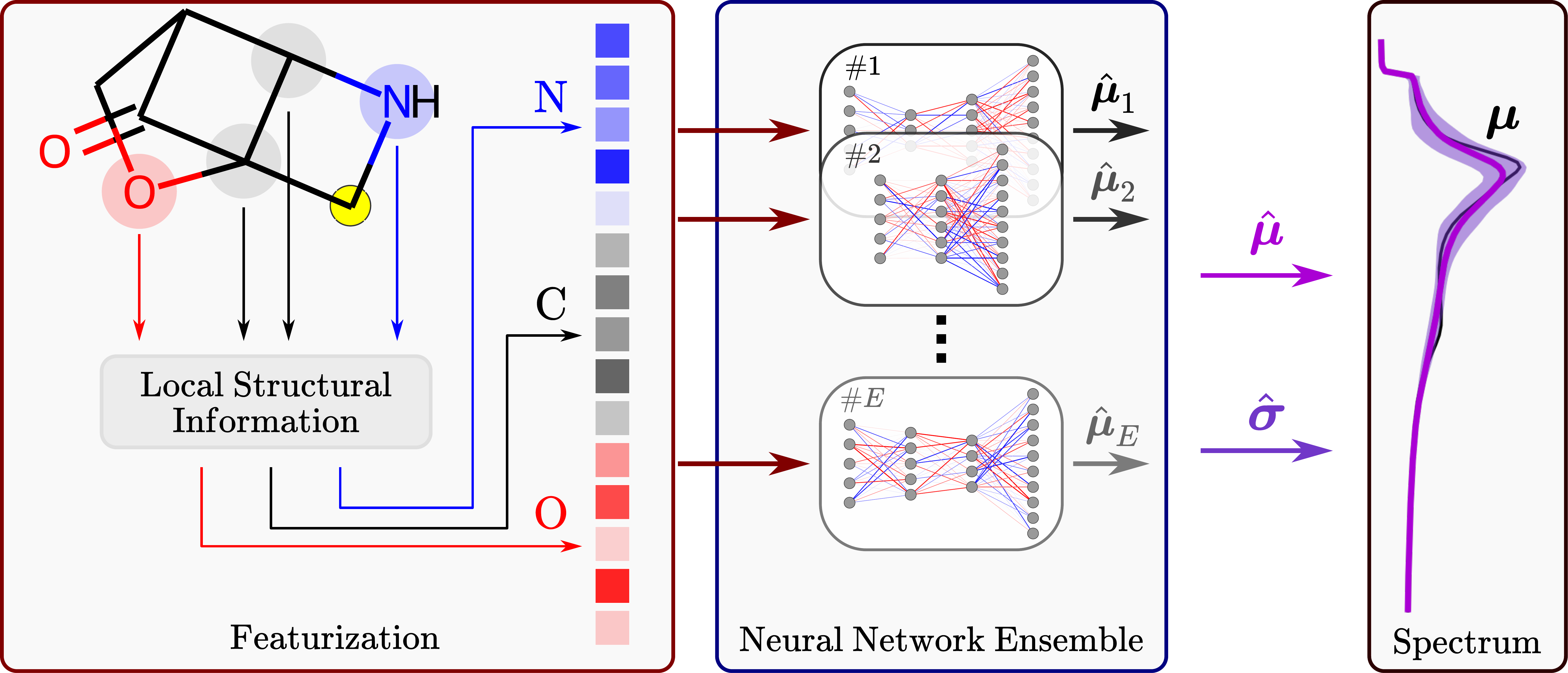}
    \caption{\label{fig:diagram} A cartoon
    of our workflow showcasing the three required steps for inference using the NNE method on a molecule in its optimized geometry. \textit{Featurization}: the atom type-resolved local structural information (red, gray and blue highlights) about the absorbing atom (yellow highlight) is encoded into a fixed-length vector representation. Fluorine and hydrogen are also encoded during featurization (for visual clarity, these are not shown in this example). \textit{Neural Network Ensemble}: each of the estimators receives the identical input vector, and each outputs a spectrum. The results are averaged over the NNE, and the spread determined. Each estimator is depicted above as having 5 inputs and 10 outputs. However, in this work, input vectors have $M=155$ elements, and the output spectra have $\Mprime=200$ elements. \textit{Spectrum}: the predicted spectra $\hat{\boldsymbol{\mu}}$ and the estimate for its uncertainty $\hat{\boldsymbol{\sigma}}$ are shown, with the ground truth $\boldsymbol{\mu}$ for reference.}
\end{figure*}

Recent years have witnessed the emergence of a thriving research enterprise directed towards the application of data-driven science to condensed matter physics, chemistry, and materials science~\cite{butler2018machine,RevModPhys.91.045002}. In particular, machine learning (ML) models such as artificial neural networks, which are universal approximators that in principle can fit any function, have been widely used to model complex relationship among physical quantities. The intersection of ML tools, emerging high-performance computing platforms and a growing number of large open source datasets has made a transformative impact on research in the physical sciences.

In the context of first-principles simulations, it has been demonstrated that ML can be used to predict molecular or materials properties from atomistic structure at comparable accuracy to the quantum mechanical theories used to produce their training data, but at only a tiny fraction of their computational cost~\cite{Behler2007,ramakrishnan2015big,wu2018moleculenet,butler2018machine}. As a result, ML has the potential to tremendously accelerate computational studies, bridge first-principles simulations to a larger time and length scales, and enable efficient materials discovery pipelines~\cite{gomez2018automatic}. Similarly, ML surrogate models can also be used to bypass the numerical solution~\cite{nomura2017restricted,PhysRevB.90.155136} and to explore the quantum states~\cite{PhysRevB.96.195145,PhysRevX.7.031038} of model Hamiltonians.

While a trained ML model provides a prediction, it usually does not provide a measure of its confidence, despite the crucial importance of model uncertainty for the researchers who apply them. In particular, ML models are designed to make accurate predictions on inputs sampled from the same distribution as the training set. However, they often fail completely when tasked with predicting on data sampled from a different distribution. Importantly, it is not always obvious (or detectable via some heuristic) when a model is performing inference on an out-of-sample input. In order to detect when this happens, one needs methodologies that incorporate uncertainty quantification (UQ). These are broadly classified as predictive methodologies that include accurate estimates of different types of statistical uncertainty. In the domain of ML and surrogate modeling, this often refers to the ability of trained models to provide some measure of confidence in the accuracy of their predictions~\cite{abdar2021review}. In research scenarios where out-of-sample data are likely to be frequently encountered, the ability of ML models to perform UQ becomes crucial.

Understanding model confidence is a key piece of the ML pipeline. Most research work simply evaluates model performance on a testing set, treating that as a proxy for understanding on-the-fly model performance. UQ takes this one step further. For example, UQ is a critical component of Gaussian Process-~\cite{Rasmussen2006GP} and neural network ensemble (NNE)-based Bayesian optimization, which has been employed for the autonomous design of experiments in many different domains, from the design of nanopaticles via flow reactors~\cite{krishnadasan2007intelligent,fitzpatrick2016novel,epps2020artificial} to the optimization of mechanical properties of materials~\cite{gongora2020bayesian,gongora2021using}. Neural network potentials (NNPs)~\cite{behler2007generalized,behler2011atom,artrith2011high,artrith2012high,schran2020committee,behler2021four,friederich2021machine} often utilize UQ to predict where their models are failing, and where they require re-training~\cite{podryabinkin2017active,sivaraman2020machine} (known as active learning~\cite{settles2009active}). Other works apply UQ in active learning for data-efficient prediction of molecular properties, such as the enthalpy, atomization energy, polarizability, and HOMO/LUMO energy levels~\cite{gubaev2018machine}. However, existing uncertainty-aware (UA) models are mostly limited to several specific topics and not broadly applied in ML applications.

In this study, we present a NNE method for quantifying the uncertainty of predicted \textit{vector} targets. Specifically, we use local atomic environment information to predict the X-ray absorption spectra (XAS) of small molecules. From existing literature on UQ implementations in ML models, very little can be discerned about their performance on spectral functions and vector targets in general, as it is unclear how the standard aforementioned applications would generalize from predicting scalars to a much higher dimensional space. We consider this XAS problem as a case study, but note that our approach is completely general. It can be applied in the broader context of \textit{any} molecular or materials property. We will show that our NNEs are not only capable of making quantitatively accurate predictions of the XAS spectra, but also of making accurate estimates of the point-wise uncertainty of said predictions.

XAS is a widely used element-specific materials characterization technique that is sensitive to the local chemical environment of the absorbing sites~\cite{ankudinov2002sensitivity,ankudinov2003development,bazin2003limits,ciatto2011evidence,ma2012defects,kuzmin2014exafs}. However, interpreting XAS data is non-trivial. While some important, physically-motivated heuristics are well known, full understanding of the relationship between spectra and underlying atomic-scale structure is mediated by electronic states. In particular, first-principles XAS simulations are playing an essential role in XAS analysis, allowing the interpretation of precise structure-property relationships otherwise much more challenging to resolve experimentally. In an XAS simulation, the spectrum is calculated from the atomic arrangement of the system. Depending on the complexity of the theory and system, the spectral simulation can be prohibitively time consuming. This limits its use for fast structure screening/refinement or spectral feature assignment. As a result, there is a growing interest to develop surrogate models that can predict XAS spectra, and other types of spectral targets, from atomistic structures~\cite{rankine2022accurate}.

% In this work, we specifically focus on C, N and O near-edge features, knows as X-ray absorption near edge structure (XANES), in molecular systems. Specifically, we focus on K-edge spectroscopy, corresponding to the excitation of a core 1s level electron to higher energy orbitals or the continuum. 

We focus on the near-edge portion of the X-ray absorption spectrum, known as X-ray absorption near edge structure (XANES). Specifically we consider K-edge XANES, corresponding to excitation from a 1s core orbital electron to empty orbitals or the continuum. We simulate the K-edge XANES spectra of C, N, and O atoms in small molecular systems for a large database of Density Functional Theory (DFT)-relaxed molecular structures: QM9~\cite{ramakrishnan2014quantum}. Taking a divide-and-conquer approach, the NNEs are trained only on the local environments of individual absorbing atoms, and molecular spectra are constructed by averaging the predictions of the individual absorption sites. Errors are interpreted as standard deviations and propagated accordingly. Fig.~\ref{fig:diagram} demonstrates the overall workflow. This allows the NNEs to make predictions on molecules larger than what the models were trained on (similar to how neural network potentials can generalize to larger systems if correlations are sufficiently short-range). As such, this approach could have broad implications in the fields of inverse design and generalizable surrogate modeling.

Key to any UQ methodology is understanding how trained models perform when data is pushed out-of-sample during inference. The addition of UQ to a surrogate model supports its generalized use. In the context of molecular systems specifically, we enumerate four physically motivated classes of generalization in which a UQ methodology can flag adverse effects of the changing local environment on model performance:
\begin{itemize}
    \item \textit{Chemical}, e.g. including larger molecules relative to the training set,
    \item \textit{Configurational}, e.g. including structural distortions from equilibrium geometry (such as due to thermal effects),
    \item \textit{Electronic}, e.g. introducing new chemical motifs (such as aromaticity) that are not necessarily a function of molecular size,
    \item \textit{Environmental}, e.g. introducing molecule-solvant interaction due to solvation.
\end{itemize}
In principle, UQ methods should be able to detect when out-of-sample data due to any of the above situations occurs. In this work, we directly study how new distributions of testing data due to chemical and configurational changes affect model and UQ performance. Studies of electronic and environmental effects are beyond the scope of the current work.

As noted, to date, incorporation of UQ into ML applications for materials science and chemistry has been limited in scope, particularly focusing on scalar target quantities (we highlight a notable exception, in which various UQ-enabled ML methods were used to predict the X-ray \textit{emission} spectra of transition metal complexes~\cite{penfold2022deep}). In the present work, with our focus on spectroscopy, we explore the challenges associated with incorporating UQ into models where the target space is of a substantially higher dimension, corresponding to the vector of spectroscopic intensity versus X-ray photon energy. In particular, this is the first time UQ methods and the requisite analysis have been demonstrated for XAS prediction.

The manuscript is organized as follows. In Section~\ref{sec:theory}, we outline the procedures used for constructing our databases, and describe the featurization and forward modeling. This includes a brief discussion on our choice of local feature embedding and various forms of UQ. Next, in Section~\ref{sec:data-analysis-prep}, we outline our procedures for analyzing the database using unsupervised and general data-analysis techniques, as well as final data preparations for machine learning. Section~\ref{sec:results} contains the main results of this work, and demonstrates the ability of our trained models to make accurate predictions and perform UQ on different subsets of small molecules. Subsection~\ref{subsec:qm9-ml-generalized} specifically highlights the power of our models to generalize to molecules with more atoms than those used during training. In Subsection~\ref{subsec:distort}, we also discuss the NNE's ability to accurately quantify uncertainty when tested on structures which are not in their relaxed geometries, a different type of generalization from the training database. Finally, we conclude and discuss the outlook and future plans of this work in Section~\ref{sec:conclusions}.

\section{Theory} \label{sec:theory}

The general theory of ensemble learning and UQ has a rich history in the ML literature. In this section, we summarize previous work and highlight the key principles and theory behind ensemble learning and UQ. Specifically, we provide an overview of supervised learning (and feed-forward neural networks), UQ, and ensembling in Subsections~\ref{subsec:ffnn}, \ref{subsec:uq}, and \ref{subsec:nne}, respectively.

\subsection{Supervised learning and feed-forward neural networks} \label{subsec:ffnn}

In supervised learning tasks, a model $f_{\boldsymbol{\theta}}$ is tasked with learning a mapping $f_{\boldsymbol{\theta}} : \mathbb{R}^M \mapsto \mathbb{R}^\Mprime,$ where $M$ is the number of features (or the length of the input vector) and $\Mprime$ is the number of targets (or the length of the target vector). This mapping can take many forms, including polynomials, random forests, non-parametric models such as Gaussian Processes, or deep learning architectures, such as neural networks. For the purposes of this discussion, we focus on parametric models, where a finite number of parameters $\boldsymbol{\theta}$ determine the form of the approximating function $f_{\boldsymbol{\theta}}.$ During the training (or fitting) process, parameters $\boldsymbol{\theta}$ are optimized so as to minimize the loss function, which is a measure of difference between the ground truth target values $\left\{\vecy^{(i)}\right\}$
and the model predictions $f_{\boldsymbol{\theta}}(\vecx^{(i)}) = \hat{\vecy}^{(i)}.$ The data on which the model is fit is referred to as the training set. Models are fine-tuned on the cross-validation set, and the final model evaluation is performed on data the model has yet to see, usually called the testing set. For an in-depth tutorial on ML techniques and proper use, we refer the reader to Refs.~\onlinecite{wang2020machine} and \onlinecite{artrith2021best}.

We use feed-forward neural networks (FFNN) to perform the supervised learning task in all results to follow. The details of FFNNs are explained in many other works (and especially in the context of spectroscopy prediction and analysis~\cite{PhysRevB.90.155136,carbone2019classification,carbone2020machine,torrisi2020random,sturm2021predicting, miles2021machine,rankine2022accurate}), and are thus not explained here. However, one key property of FFNNs is that they are \emph{universal approximators}, meaning they can, in principle, model any function provided the model has enough trainable parameters and is trained on enough data. This is relevant for constructing ensembles of neural networks, described in Subsection~\ref{subsec:nne}. The details of our features and targets are explained in Subsection~\ref{subsec:qm9-data-prep}.

\subsection{Uncertainty quantification} \label{subsec:uq}
Generally, ML models are tasked with modeling inputs to outputs, as described in the previous subsection. However, there are currently significant ongoing efforts to leverage statistical principles to model, or quantify, the uncertainty in these predictions. For example, Gaussian Processes (GPs)~\cite{Rasmussen2006GP} are non-parametric generalizations of the multivariate normal distribution to the continuum, and are finding widespread use due to their ability to rigorously quantify statistical uncertainty. This uncertainty is derived from assumptions about correlation lengths embedded in a covariance kernel, allowing one to draw samples from the GP consistent with the parameters of the embedded length scales. For the purposes of this work, we henceforth outline ways to quantify uncertainty in \textit{parametric} models such as FFNNs. We reserve the discussion of ensembling specifically to Subsection~\ref{subsec:nne}.

Uncertainty-aware models incorporate UQ in order to address two different types of uncertainty: aleatoric and epistemic~\cite{hullermeier2021aleatoric,abdar2021review}. Aleatoric uncertainty is also called ``irreducible," as it is due to natural physical processes (such as randomness in nature) or inherent instrument error. Intrinsic broadening processes in spectroscopy or noise in an image are two examples of aleatoric uncertainty. We highlight that error bars corresponding to uncertainty in a physical measurement are also examples of aleatoric uncertainty. Epistemic uncertainty is due to an insufficient model. It is often large when training data in some region of the input space is not adequately sampled. Unlike aleatoric uncertainty, epistemic uncertainty can be improved by using some combination of a more sophisticated model, incorporating more training data or prior information. While it is non-trivial, recent work has demonstrated that both classes of uncertainties can be predicted using ML techniques~\cite{egele2021autodeuq}.

One method for modeling aleatoric uncertainty in particular is the mean-variance estimation (MVE)~\cite{nix1994estimating}. A MVE model will attempt not only to predict the target value, but also an estimation for the uncertainty in that prediction as another output. Concretely, given an input feature dataset $X \in \mathbb{R}^{N \times M}$ (where $N$ is the number of training examples) and a target dataset $Y \in \mathbb{R}^{N \times \Mprime},$ a MVE model will attempt to learn a mapping $f : \mathbb{R}^{M} \mapsto \mathbb{R}^{2\Mprime}.$ The output space is doubled in size since for every output prediction, an uncertainty estimate is also predicted. If we consider the scalar output case $\Mprime=1,$ we have a single prediction $\hat{y}$ and a single prediction for the estimate of the uncertainty $\hat{\sigma}.$ The MVE model is trained to minimize a negative Gaussian log-likelihood (NLL) loss function,
\begin{equation} \label{eqn:nnl}
    L(y, \hat{y}, \hat{\sigma}) = \frac{1}{2} \log 2\pi + \frac{1}{2}\log \hat{\sigma}^2 + \frac{\left( y - \hat{y} \right)^2 }{2 \hat{\sigma}^2}.
\end{equation}
The principle is simple: if the model makes an accurate prediction (i.e. $(y - \hat{y})^2$ is small), it can ``afford" to also predict a low uncertainty $\sigma.$ However, if the prediction error is large and cannot be improved during training (perhaps due to a significantly noisy observation) the model will compensate by increasing $\hat{\sigma},$ despite paying the penalty of the second term in Eq.~\eqref{eqn:nnl}. The fact that a MVE model trained using the NLL loss function has the flexibility to make this tradeoff allows it to estimate the uncertainty in its own predictions, offering considerable utility when modeling irreducible noise.

In our work, the simulation of a XANES spectrum from molecular coordinates is deterministic. Hence, we will only be concerned with modeling epistemic uncertainty. There are several uncertainty quantification techniques that can model epistemic uncertainty, such as Bayesian neural networks (BNN)~\cite{jospin2022hands}, Monte Carlo dropout (MCD) ensembles~\cite{srivastava2014dropout} and neural network ensembles~\cite{hu2019mbpep,wilson2020bayesian}. Instead of directly predicting an estimate of uncertainty, BNN's treat all of their parameters as random variables, allowing one to sample from this distribution during inference, leading to a distribution of predictions. While proven to be extremely effective in certain cases~\cite{tschannen2018recent,zhu2019physics,luo2022bayesian}, BNN's are expensive to train, and it has been argued that they are consistent with simpler, ensemble-based approaches~\cite{wilson2020bayesian}. MCD ensembles work under a similar principle, by randomly disabling neurons during inference, allowing for a sampling over many ``effective models" and thus allowing ensemble-like predictions to be obtained. Statistical bootstrapping can also be used to generate model diversity, and has been shown to be superior to MCD in similar applications~\cite{penfold2022deep}. In this study, we choose to use NNEs for the reasons explained in detail below. Similar to Ref.~\onlinecite{penfold2022deep}, we also downsample our training set size in a way similar to bootstrapping, and combine this with ensembling to create even more model diversity than either would produce on its own (and also found that MCD is less effective than our combined downsampling and ensembling method).

\subsection{Neural network ensembles} \label{subsec:nne}

NNEs are sets of individual models (or estimators, in the case of this work these are FFNNs) which are usually trained independently but used together during inference (see Fig.~\ref{fig:diagram}). The working principle of NNEs is that of the wisdom of the crowd (or ``query-by-committee"~\cite{settles2009active}): each individual's prediction is less robust than the aggregate opinion. More rigorously, any two models that produce an accurate result will \textit{by definition} produce predictions that are similar. However, due to the vast training weight-space of deep learning models, when any two models both fail, they will usually fail in different ways. The training weight-space refers to the complete set of trainable parameters of a neural network. The values of these weights are randomly initialized between different estimators, and since this space can be incredibly high-dimensional (easily $>1$ million and often many orders of magnitude larger), there are a vast number of possible sets of weights corresponding to local minima in the landscape of the loss function. The ensemble learning approach turns one of the neural network's greatest weaknesses into a strength. Each estimator (in its own local minimum) will produce roughly the same correct prediction in a well trained neural network and is partly what allows neural networks to be so flexible. However, the differences between these sets is also what leads to different estimator predictions in failure scenarios.

Empirically, the average over the entire ensemble not only produces better predictive accuracy, but also allows for UQ by interpreting the spread in the predictions of the individual estimators. Theoretically, an ensemble-averaged prediction can be thought of as an averaging over the space of ``reasonable" possible functions mapping inputs to outputs given a fixed training set. This space is infinite, of course, and any finite sampling of models is not sufficient to rigorously cover this space. However, it is sufficient to provide useful uncertainty measures. Choosing how many estimators to use remains a highly problem- and model-dependent open research question~\cite{settles2009active}. This is in stark contrast to a GP, which not only provides an analytic form for the mean and spread of the GP averaged over an infinite number of estimators, the spread itself is rigorously the standard deviation of a Gaussian distribution. A NNE may not adequately approximate the space and the spread is not strictly interpretable as a proper standard deviation (though it can be used in a similar way~\cite{schran2020committee}). We thus highlight a critical pitfall: like almost all ML techniques, especially in deep learning, UQ is hyperparameter-dependent. These methods must be rigorously tested in order to ensure they are of the appropriate quality given the problem at hand. To our knowledge, there is not yet any formal theory for interpreting the distribution of predictions of NNEs.

In this study, we choose NNEs for their balance of relative simplicity, predictive power and overall performance. We also highlight that they operate on essentially the same paradigm that any individual estimator does, which makes them straightforward to train, debug and deploy.

\section{Methodology} \label{sec:data-analysis-prep}

In this work, we showcase the utility of the NNE method for UQ on molecular structure-XAS pairings. Molecular structures are taken from the QM9 database~\cite{ramakrishnan2014quantum}, which is a subset of the GDB-17 chemical universe~\cite{ruddigkeit2012enumeration}. QM9 contains roughly 134k DFT-geometry optimized small molecules, each with at most 9 heavy atoms (C, N, O, F). Molecular spectra are computed using the multiple scattering code FEFF9~\cite{rehr2010parameter}. We focus on C, N and O K-edge XANES spectra from individual absorbing sites, and as such we partition our database into $\mathcal{D}_A$ for $A = \{\mathrm{C}, \mathrm{N}, \mathrm{O}\}.$ For example, $\mathcal{D}_\mathrm{O}$ is a database containing all oxygen site-XANES pairs. In the following subsections, we explain how our features and targets are constructed (Subsections~\ref{subsec:feature-construction} and \ref{subsec:target-constrction}, respectively) and analyzed (Subsection~\ref{subsec:QM9-pca}). Finally in Subsections~\ref{subsec:qm9-data-prep} and \ref{subsec:methods-ML}, we describe the procedure for setting up the training and testing sets used in the remainder of the work, and implementing our NNE approach.

\subsection{Feature construction} \label{subsec:feature-construction}

XANES is sensitive to the local chemical environment of absorbing atoms. We therefore choose a structural descriptor that is local to each absorbing site: Atom-centered Symmetry Functions (ACSFs)~\cite{behler2007generalized,behler2011atom}. We also considered other descriptors (particularly those from the Dscribe library~\cite{dscribe}) including the Smooth Overlap of Atomic Positions~\cite{bartok2013representing} and Many-body Tensor Representations~\cite{huo2017unified}, but ultimately decided to use ACSF.\footnote{We also considered the weighted-ACSF (wACSF) feature encoding~\cite{gastegger2018wacsf}, which unlike the traditional ACSF, do not scale in size with the number of unique atom types in the considered data. While this type of encoding is particularly useful when there are a significant number of unique atom types (such as when dealing with the vast spaces of materials or materials complexes~\cite{carbone2019classification,penfold2022deep}), it is not necessary for our problem, as we only consider 5 unique atom types (H, C, N, O and F).} The ACSF feature encodings have been very successful in modeling total energy partitioned into local atomic contributions. ACSFs were first proposed by Behler in 2011 in the context of developing neural network potentials~\cite{behler2007generalized,behler2011atom} (NNPs). The development of NNPs is summarized in a recent review by Behler~\cite{behler2021four}, along with the utility and flexibility offered by ACSFs.

ACSF feature vectors are further described at length in quite a few recent works, including Refs.~\onlinecite{artrith2016implementation,rankine2022accurate}, and as such will not be repeated here. In brief, the ACSF feature vectors are atom-resolved representations of the local radial and angular atomic environments of a central atom, which in this work is the absorbing site.

\begin{figure}[!htb]
    \centering
    \includegraphics[width=\columnwidth]{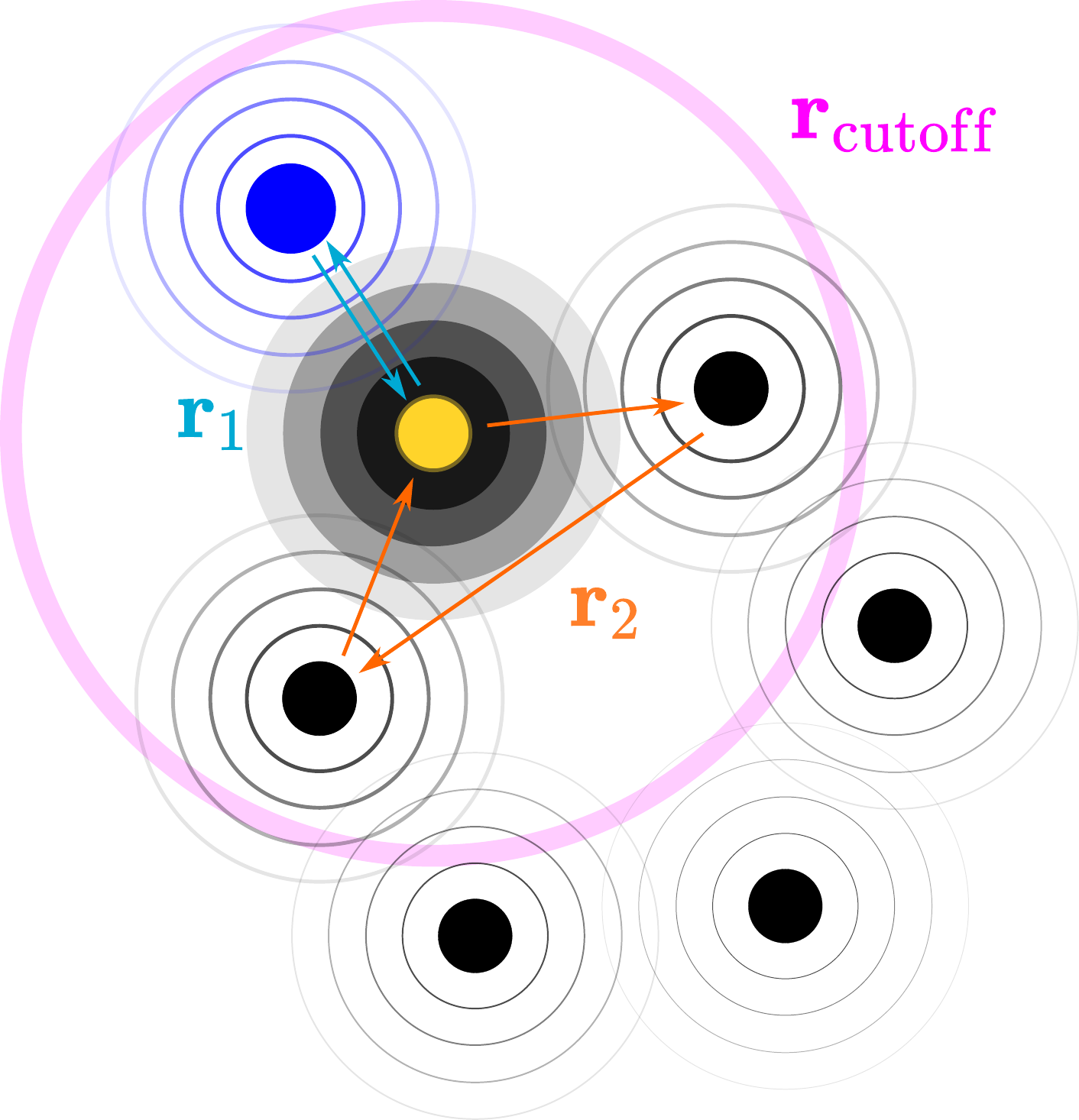}
    \caption{\label{fig:scattering} A cartoon depicting two scattering paths from a central absorbing atom (marked in yellow; other C atoms are black) in an Aniline molecule (with hydrogen atoms implicit, and N is blue). The first path is C$\rightarrow$ N$\rightarrow$ C, and the second C$\rightarrow$ C$\rightarrow$ C $\rightarrow$ C. The cutoff region around the central atom marks the geometric information included in the ACSF feature vectors. Information outside of this cutoff (which corresponds to relatively long scattering path lengths) is excluded.}
\end{figure}

We aim to leverage the locality of XANES in the same way local atomic energy contributions are in constructing NNPs. Physically, the argument that XANES is a local probe can be understood through multiple scattering theory, where the absorption coefficient at a given incident photon energy is determined by the interference between the outgoing wave and the back scattering waves from neighboring atoms. In the multiple scattering path expansion, the absorption coefficient decays exponentially with path length~\cite{ankudinov1998real,rehr2000theoretical}. Overall, the longer the scattering path length, the smaller the overall contribution to the XANES spectrum. Typical paths that contribute are illustrated in Fig.~\ref{fig:scattering}, where $\vecr_1$ and $\vecr_2$ represent two- and three-atom scattering paths, respectively. Of course, a much larger number of paths contribute in principle, but longer paths, mostly those outside of the range $\vecr_\mathrm{cutoff},$ do not contribute significantly.

In preparing our ACSF features, for every absorbing atom site, we use a radial cutoff of 6~\AA, as well as similar parameters to those used in Ref.~\onlinecite{artrith2016implementation}. H, C, N, O and F neighbors were considered, and for each absorbing atom site, a feature vector of 155 entries was constructed. The details of our featurization process can be found in Appendix~\ref{appendix:acsf}.

We explored feature reduction/importance ranking, similar to that done in Ref.~\onlinecite{rankine2022accurate}, in order to reduce the 155-dimensional input vector to a smaller dimension. We found that training was somewhat stabilized (i.e., loss functions decreased monotonically more consistently), but accuracy overall was not noticeably different from training performed using the full ACSF vector. Thus, we choose to use the ACSF feature vectors as-is for inputs to our models.

\subsection{Target construction} \label{subsec:target-constrction}

As previously mentioned, FEFF9~\cite{rehr2010parameter} is used to compute the XANES spectra using multiple scattering theory. Each molecule's FEFF spectrum is computed individually (atom-by-atom), and the details of these calculations are presented in Appendix~\ref{appendix:spectra}. In brief, we use a cutoff of 7~\AA~for self-consistent potential calculations and 9~\AA~for full multiple scattering calculations, which is commensurate with the geometric cutoff of 6~\AA~used for the ACSF descriptor construction. The targets are the XANES spectra interpolated using cubic splines onto a common grid, which was chosen to be 200 dimensional, corresponding to a resolution of 0.27 eV. 

In brief, the spectral target can be represented as a vector
\begin{equation} \label{eq-mu-vector-ground-truth}
    \boldsymbol\mu^{(i)} = \left[\mu_1^{(i)}, \mu_2^{(i)}, ..., \mu_M^{(i)}\right]
\end{equation}
for training example $i,$ where $M=200$ is the number of target values. We consider a spectral range of 50 eV and scale the intensity to unity at the high energy tail, conforming to the standard XANES normalization procedure.

\subsection{Principal Component Analysis} \label{subsec:QM9-pca}

\begin{figure*}[!htb]
    \centering
    \includegraphics[width=2\columnwidth]{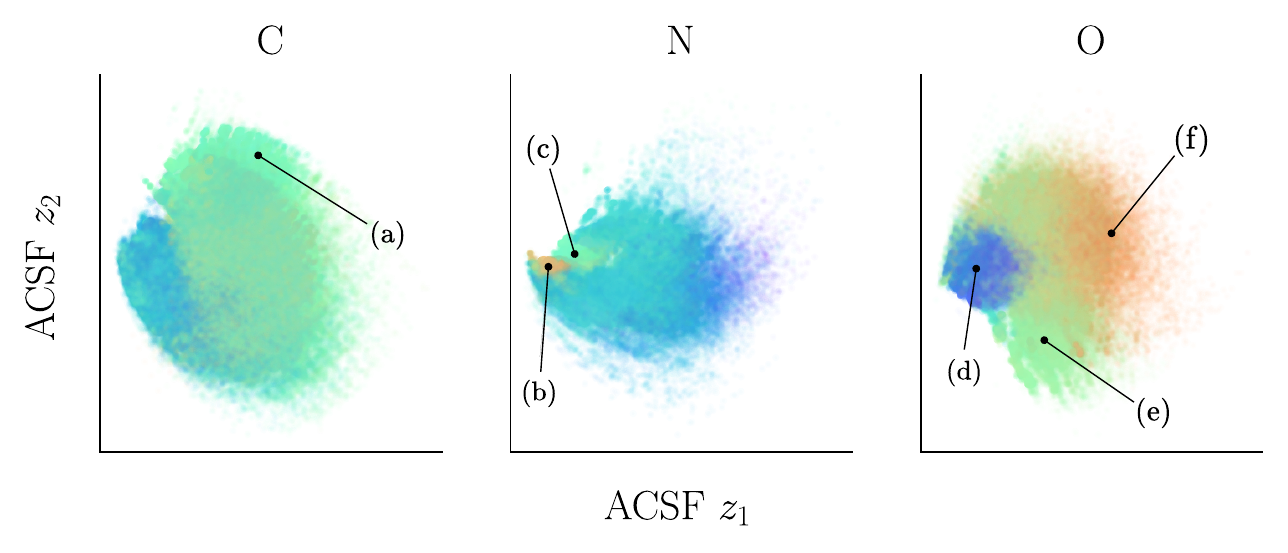}
    \caption{\label{fig:qm9_pca} Principal component analysis of the feature(structure)-target (spectrum) relationship for each of the three datasets of C, N and O absorbing atoms. $z_1$ and $z_2$ are weights along the first and second principal directions ($\vecw_1, \vecw_2$) as constructed on the ACSF feature data. The color of the markers correspond to the first principal value of the spectral target data. Regions of significant color density are indicated by labels (a)-(f).}
\end{figure*}

Prior to performing ML modeling, it is always prudent to explore the data to ensure sensible correlations or patterns exist between features and targets. It also provides the baseline intuition of what to expect from the ML models. Linear dimensionality reduction techniques, such as Principal Component Analysis (PCA), cannot capture the non-linear relations that a neural network will, but they are still quite useful in identifying overall trends and are largely parameter-independent. More sophisticated non-linear techniques, such as t-Distributed Stochastic Neighbor Embedding (t-SNE)~\cite{van2008visualizing}, can extract more complicated trends, but are usually highly parameter dependent and thus less robust~\cite{wattenberg2016use}. We therefore apply PCA on both the ACSF features and spectra targets in order to resolve their relations in a tightly controlled manner.

PCA extracts the ``directions of principal variance" in a dataset. The PCA decomposition diagonalizes the $M \times M$ covariance matrix of a dataset $X \in \mathbb{R}^{N \times M}$ ($N$ examples each with $M$ features). The most significant eigenvectors and eigenvalues (the eigenvectors which correspond to directions of maximal variance are indicated by the largest eigenvalues) of the covariance matrix are used to project the data into a lower-dimensional space. Formally, the (scaled) eigenvalues $\mathbf{w}_j$ are the (relative) captured variance $\omega_j$ along the direction defined by that eigenvector. For the $i$th example in the database $\mathbf{X}_i$, and for the $j$th eigenvector $\mathbf{w}_j$,
\begin{equation}
    z_{ij} = \mathbf{X}_i \cdot \mathbf{w}_j,
\end{equation}
where $\cdot$ is the dot product. For example, $z_{i1}$ captures the first principal component (in the direction of maximal variance) of example $i.$ We also highlight that given a value $\vecz_i = [z_{i1}, z_{i2}, ..., z_{id}]$ with $d < M,$ an approximate reconstruction of $\mathbf{X}_i$ can be obtained via
\begin{equation}
    \mathbf{X}_i \approx \sum_{j=1}^d z_{ij} \mathbf{w}_{j}.
\end{equation}

\begin{table}[!htb]
\caption{\label{tab:captured variances}%
The relative variance captured by the first two principal components of both the ACSF and spectral ($\mu$) spaces.
}
\begin{ruledtabular}
\begin{tabular}{lcccc}
\textrm{Absorber}&
\textrm{$\omega_1$(ACSF)}&
\textrm{$\omega_2$(ACSF)}&
\textrm{$\omega_1$($\mu$)}&
\textrm{$\omega_2$($\mu$)} \\
\colrule
\text{C} & 0.61 & 0.16 & 0.34 & 0.29 \\
\text{N} & 0.74 & 0.09 & 0.41 & 0.26 \\
\text{O} & 0.77 & 0.11 & 0.60 & 0.22
\end{tabular}
\end{ruledtabular}
\end{table}

Using the scikit-learn library~\cite{scikit-learn}, we apply PCA to the feature (the ACSF vectors) and target (spectra) spaces, independently. Specifically, we perform the dimensionality reduction on the ACSF feature data as $\mathbb{R}^{N \times 155} \rightarrow \mathbb{R}^{N \times 2}$ and the spectra target data as $\mathbb{R}^{N \times 200} \rightarrow \mathbb{R}^{N \times 2}$. The values of $z_{i1}$ and $z_{i2}$ for the ACSF decomposition are plotted in Fig.~\ref{fig:qm9_pca} on the $x$ and $y$ axes, respectively. The color value of the points represents the value of $z_{i1}$ of the \emph{spectra} decomposition. Note that scales are not shown here, as they not important for the qualitative analysis to follow. Table~\ref{tab:captured variances} tabulates the relative captured variance of each dimension for both the features and targets, in the first two principal directions.

Analysis of Fig.~\ref{fig:qm9_pca} shows clear spatial correlations between the principal values of the ACSF features and spectral targets. Areas of high color density indicate a spectral feature which is strongly correlated to a common structural motif. These regions are indicated by the appropriate labels. For example, while the C atom clustering is the most poorly resolved, cluster (a) appears to correspond to aliphatic carbon chains. For N atoms, (b) clearly corresponds to azides and (c) to primary ketimines. For O atoms, (d), (e) and (f) correspond to esters, alcohols and ethers, respectively. While the clustering patterns are not definitive on their own, they indicate a high degree of correlation between the XANES spectra and functional group, a result observed in previous work~\cite{carbone2020machine}. Not only does this further substantiate the locality of XANES, it also hints that ML techniques will be able to efficiently capture a more complicated non-linear relationship between XANES spectra and local atomistic geometry.

\subsection{Data splits and preparation} \label{subsec:qm9-data-prep}
We test two hypotheses using the QM9 dataset, each necessitating different partitionings of the datasets $\mathcal{D}_A.$ First, the ACSF feature vectors capture sufficient local structural information about absorbing atoms for accurate prediction of site-wise XANES spectra and uncertainty estimations. Testing our first hypothesis involves evaluating the overall effectiveness of a NNE trained using the usual \textit{random} train/validation/test split. It corresponds to a use case in which the trained NNE is expected to perform on a randomly selected example in the QM9 database. Hence, this first partitioning is referred to as the ``random partitioning" ($\mathcal{D}_A^\mathrm{R}$). Such a performance measure is perfectly valid if the distribution of molecules in the test set is chemically similar to those found in the QM9 training set.

The second hypothesis is that the XANES spectra are sufficiently local such that an ensemble can be trained on data containing molecules with fewer than $n$ heavy atoms, but still perform on molecules with more than $n$ heavy atoms. Furthermore, the individual local signals can then be averaged, with NNE error propagated, to estimate the molecular XANES, and its error. This is indeed a significant challenge. It is known that neural network potentials, which often also use ACSF input vectors with similar parameters, can struggle in systems with long-range correlations. This hypothesis therefore provides a stringent test on both the locality of the XANES spectra and NNE. If non-trivial long-range correlation effects exist in the XANES spectra, our models will suffer due to the intrinsic locality constraint. Similarly, if the chemical environments captured in QM9 are significantly different than those contained in a dataset with larger molecules, the NNE will fail to generalize. As this partitioning will test the ability of the models to generalize, it is henceforth referred to as the ``generalization test" ($\mathcal{D}_A^\mathrm{G}$).\footnote{Note that the same data/examples are contained in $\mathcal{D}_A,$ $\mathcal{D}_A^\mathrm{R}$ and $\mathcal{D}_A^\mathrm{G}.$ We distinguish between them to highlight that the train/validation/test splits \textit{are} different.} 

For each $\mathcal{D}_A^\mathrm{R}$, a simple random split is employed, where 90\% of data are chosen for training and cross-validation, with the rest held-out as the testing set. For $\mathcal{D}_A^\mathrm{G},$ multiple splits are made. The testing sets always contain all of the QM9 molecules with a total of 9 heavy atoms. Training sets are constructed by choosing molecules with anywhere from 5-8 heavy atoms. For example, in one training set instance for nitrogen absorbing atoms, we include sites from all molecules containing at least one N atom, but less than, e.g., total 6 heavy atoms in the training (and cross-validation) splits. Evaluation is then consistently performed on atoms originating from molecules with 9 heavy atoms.

Significantly, the data partition $\mathcal{D}_A^\mathrm{G}$ has a strong impact on the size of the training set. Binned by the total number of heavy atoms per molecule, the total number of molecules increases exponentially in the QM9 database.
\begin{figure}[!htb]
    \centering
    \includegraphics[width=\columnwidth]{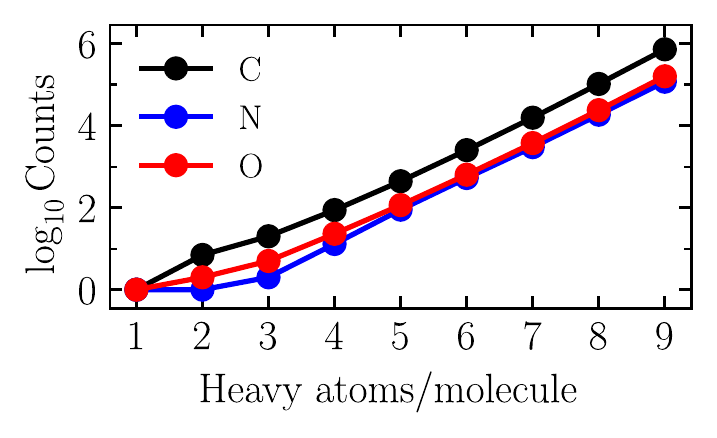}
    \caption{\label{fig:qm9_N} The total number of molecules contained in QM9 as a function of the number of heavy atoms (C, N, O and F)/molecule for each of the three datasets considered in this work. Each dataset, $\mathcal{D}_A$ (labeled by $A$ for brevity) corresponds to the subset of QM9 in which each molecule contains at least one atom of type $A.$}
\end{figure}
In Fig.~\ref{fig:qm9_N}, we show the total number of molecules in $\mathcal{D}_A$ as a function of the number of heavy atoms/molecule. As one can see, the total amount of data for 9 heavy atoms/molecule is roughly an order of magnitude greater than for 8. If the generalization test succeeds, an exponential increase in training data could be avoided.

\subsection{Machine learning} \label{subsec:methods-ML}

We train $\abs{\mathcal{E}} = 30$ independent estimators for all experiments in this work. The details of the training procedures are given in Appendix~\ref{appendix:training-details}, and we highlight the following important points. First, each estimator was always trained on a random 90\% sampling (without replacement) of the training set~\cite{efron1994introduction,bakker2003clustering}, meaning some data was purposely excluded during training (the dependence on the sampling proportion is analyzed in Appendix~\ref{appendix:train-set-dependence}). Second, each estimator used a randomly initialized neural network architecture. Both of these procedures were employed to maximize model diversity, which as previously discussed, has been shown to be of great utility for UQ. 

During initial cross-validation studies, we observed that occasionally individual estimators would produce completely unphysical results. Such results include, but are not limited to, spikes in the spectral intensity an order of magnitude larger than the most intense spectrum in our data and ``vanishing" spectra with mostly zero intensity. Likely due to a combination of the random model initialization and training set downsampling, these aberrant results do not contribute meaningful information to either the overall accuracy of the prediction or the uncertainty estimate. Therefore, during inference, these faulty prediction-estimator pairs are discarded when computing ensemble-averaged quantities. We discuss the details of this procedure in Appendix~\ref{appendix:ensemble-prediction-details}.

\section{Results}\label{sec:results}

\subsection{Random partitioning} \label{subsec:qm9-ml-random}

The NNE predictions for spectrum $i$ on spectral grid point $j$ is given by an average over the individual estimators,
\begin{equation}
    \hat{\mu}_j^{(i)} = \frac{1}{\abs{\mathcal{E}(i)}} \sum_{k \in \mathcal{E}(i)} \hat{\mu}_j^{(i, k)},
\end{equation}
where $k$ is the estimator index, and $\mathcal{E}(i) \subseteq \mathcal{E}$ is the set of estimator indexes corresponding to non-outlier, physical predictions for example $i$ (and $\abs{\mathcal{E}(i)}$ is the size of this set and $\mathcal{E}$ is the set of all estimators). The ensemble-averaged error is
\begin{equation} \label{eqn:delta-i}
    \varepsilon^{(i)} = \frac{1}{M}\sum_{j=1}^M \varepsilon_j^{(i)},
\end{equation}
where
\begin{equation} \label{eqn:delta-ij}
    \varepsilon_j^{(i)} = \abs{\mu_j^{(i)} - \hat{\mu}_j^{(i)}}.
\end{equation}
We note that the vector representation of the predicted XANES spectrum is given in a similar form to Eq.~\eqref{eq-mu-vector-ground-truth},
\begin{equation} \label{eq-mu-vector-pred}
    \boldsymbol{\hat{\mu}}^{(i)} = \left[\hat{\mu}_1^{(i)}, \hat{\mu}_2^{(i)}, ..., \hat{\mu}_M^{(i)}\right].
\end{equation}

During inference, we use the ensemble-averaged quantity as the overall ensemble prediction. In order to demonstrate the NNE's superiority in raw predictive accuracy, we compare the ensemble prediction above to that of the average prediction error of each individual estimator,
\begin{equation} \label{eqn:delta-i-est}
    \varepsilon_\mathrm{est}^{(i)} = \frac{1}{\abs{\mathcal{E}} M} \sum_{k=1}^\abs{\mathcal{E}} \sum_{j=1}^M \abs{\mu_j^{(i)} - \hat{\mu}_j^{(i, k)}}.
\end{equation}
Equation~\eqref{eqn:delta-i-est} can be best thought of as a rough measure of how any \textit{single} model would perform on average. To quantify this, we define the average test error on a logarithmic scale over $N_\mathrm{test}$ structure-spectrum pairs,
\begin{subequations} \label{eq:epsilon-bar}
    \begin{equation}
        \bar{\varepsilon}=\frac{1}{N_\mathrm{test}}\sum_i \log_{10} \varepsilon^{(i)},
    \end{equation}
and
    \begin{equation}
        \bar{\varepsilon}_\mathrm{est}=\frac{1}{N_\mathrm{test}}\sum_i \log_{10} \varepsilon_\mathrm{est}^{(i)}.
    \end{equation}
\end{subequations}
We highlight that $\bar{\varepsilon}$ ($-1.45, -1.40$ and $-1.55$) clearly outperforms $\bar{\varepsilon}_\mathrm{est}$ ($-1.36, -1.32$ and $-1.46$ ) for the C, N and O datasets. The details of the distributions of these errors are discussed in Appendix~\ref{appendix:train-set-dependence} (Fig.~\ref{fig:qm9-all-hists}).

\begin{figure}[!htb]
    \centering
    \includegraphics[width=\columnwidth]{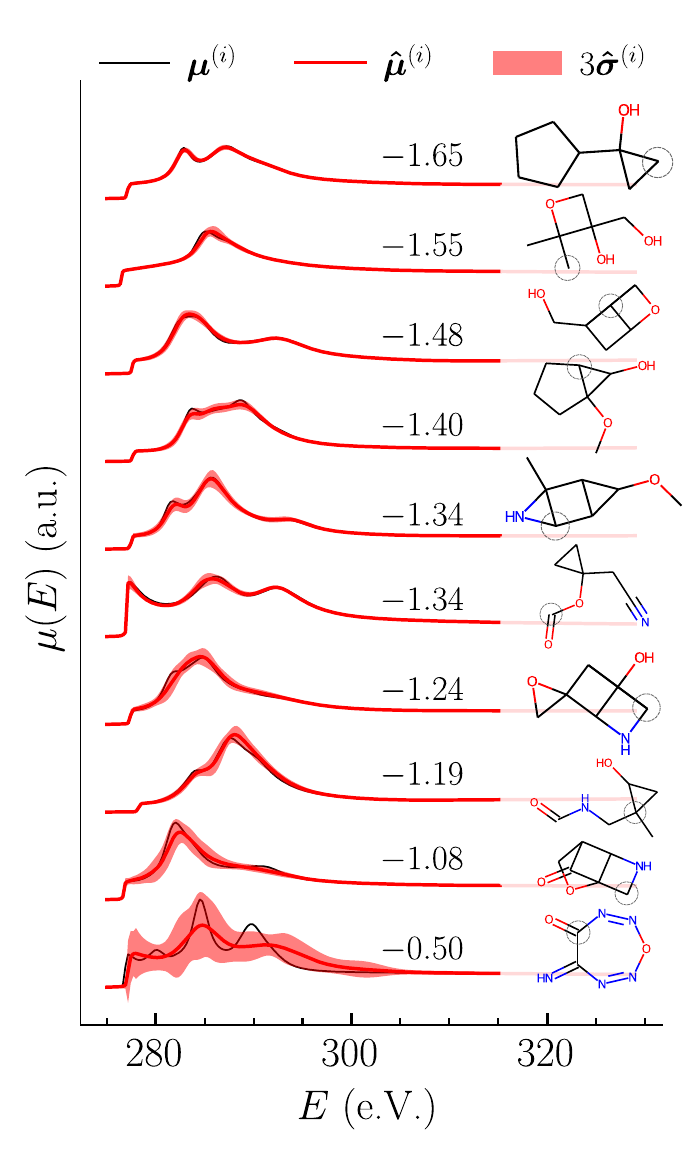}
    \caption{\label{fig:qm9_C_random_preds} Waterfall plot sampled from the bottom (worst) of each decile of the testing set results on $\mathcal{D}_\mathrm{C}^\mathrm{R},$ where decile are sorted from best (top) to worst (bottom). Absorbing carbon atom sites are indicated by a dashed circle in the molecular diagram. The ground truth (black), prediction (red) and $3\times$ the spread (shaded red) are displayed. The value for $\log_{10} \varepsilon^{(i)}$ is also shown. The vector predictions and uncertainties are given by Eqs.~\eqref{eq-mu-vector-pred} and \eqref{eqn:boldsymbol-sigma-i}.}
\end{figure}

In order to ground the discussion of overall model performance, we present waterfall plots in Fig.~\ref{fig:qm9_C_random_preds} with samples randomly chosen from the worst cases in each decile of the the testing set of $\mathcal{D}_\mathrm{C}^\mathrm{R}$. One of the worst performers (bottom figure) clearly originates from a rare, challenging (from the electronic structure perspective) structure: a 7-membered fully conjugated ring containing 5 heteroatoms. Given the chemical space covered in QM9, it is not unsurprising that the prediction is not accurate. However, it appears that the uncertainty estimate yields the qualitatively correct trend, as the prediction appropriately presents with relatively large error bars. On the other hand, all other predictions are relatively accurate, and present with error bars roughly commensurate with the prediction accuracy.

The point-wise NNE spread for spectrum $i$ is defined as
\begin{equation} \label{eqn:sigma-hat-ij}
    \hat\sigma_j^{(i)} = \sqrt{\frac{1}{\abs{\mathcal{E}(i)} } \sum_{k \in \mathcal{E}(i)} \left(\hat{\mu}_j^{(i)} -  \hat{\mu}_j^{(i, k)}\right)^2},
\end{equation}
from which the overall uncertainty of the prediction can be computed,
\begin{equation} \label{eqn:sigma-hat-i}
    \hat\sigma^{(i)} = \frac{1}{M} \sum_{j=1}^M \hat\sigma_j^{(i)}.
\end{equation}
Similar to Eqs.~\eqref{eq-mu-vector-ground-truth} and \eqref{eq-mu-vector-pred}, the vector-uncertainty for a single spectrum can be represented as
\begin{equation} \label{eqn:boldsymbol-sigma-i}
    \hat{\boldsymbol{\sigma}}^{(i)} = \left[\hat\sigma_1^{(i)}, \hat\sigma_2^{(i)}, ..., \hat\sigma_M^{(i)}\right].
\end{equation}
It is important to note that Eqs.~\eqref{eqn:sigma-hat-ij}, \eqref{eqn:sigma-hat-i} and \eqref{eqn:boldsymbol-sigma-i} are independent of the ground truth predictions. Additionally, we note that unlike e.g. a GP, the spreads $\sigma_j^{(i)}$ are \textit{not} proper standard deviations, since the distribution of estimator outputs is not guaranteed to be Gaussian. They are simply a measure of how different each output is from the others.

\begin{figure*}[!htb]
    \centering
    \includegraphics[width=2\columnwidth]{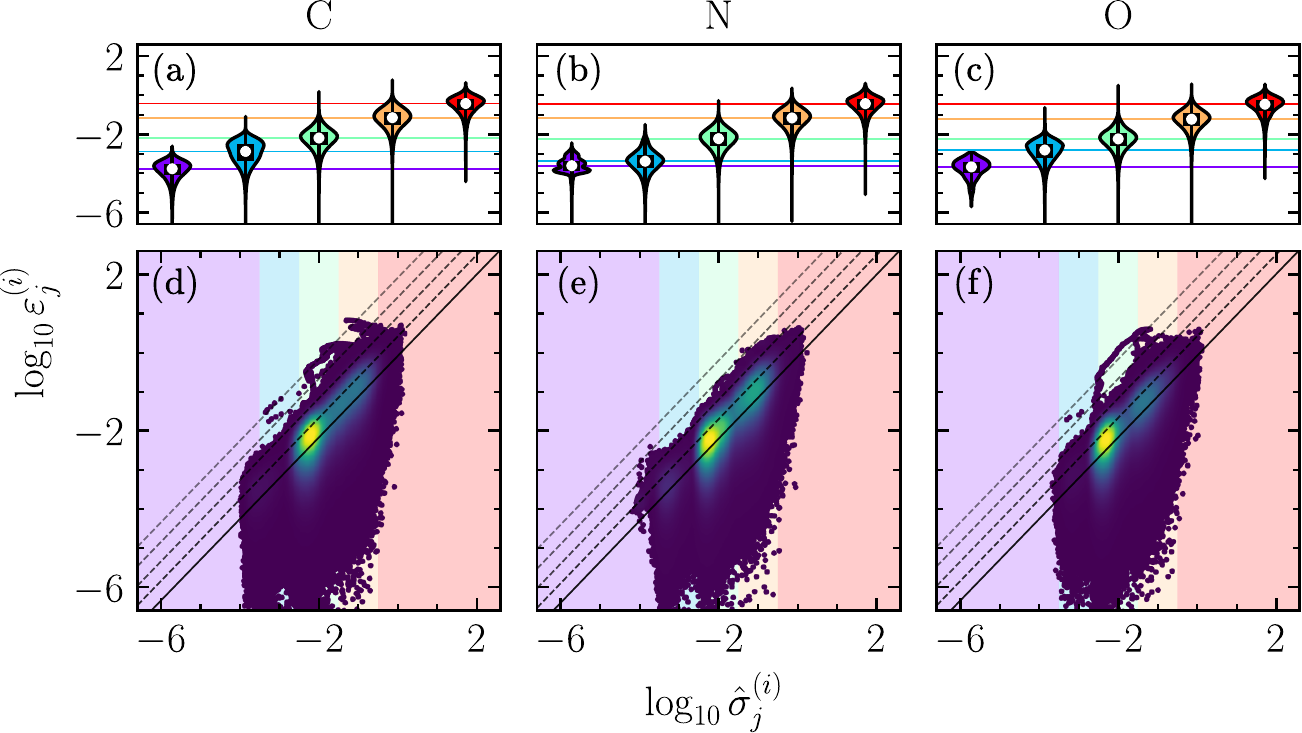}
    \caption{\label{fig:qm9-parity} Violin plots of the $\log_{10}$-ensemble error as a function of the binned $\log_{10}$ spreads of the ensemble prediction (a)-(c), and parity plots (2D density histograms) of the point-wise $\log_{10}$-ensemble error as a function of the $\log_{10}$-estimator spread (d)-(f). Violin plot colors are coordinated with the bins in the scatterplots, which are given by $\{(-\infty, -3.5), [-3.5, -2.5], [-2.5, -1.5), [-1.5, -0.5), [-0.5, \infty)\},$ and are colored purple, blue, green, orange and red, respectively; the medians of each of the bins are shown for reference as horizontal solid lines. In the scatterplots, regions of high (low) density are shown in yellow (purple). The best linear fit to the scatterplot data (solid line) is shown in addition to four guidelines at half orders of magnitude intervals above it (dashed lines). The origins of the systemic outliers in (d) and (f), those falling about the third and fourth dashed guidelines, are discussed in Appendix~\ref{apdx:outliers}, and do not have a meaningful effect on the presented analysis.
    }
\end{figure*}

We present a quantitative analysis of the NNE's capability of accurately capturing uncertainty measures in Fig.~\ref{fig:qm9-parity} for all three datasets $\mathcal{D}_A^\mathrm{R}.$ In (a)-(c), violin plots of $\log_{10}\varepsilon_j^{(i)}$ for 5 bins of $\log_{10}\hat{\sigma}_j^{(i)}$ (shown as the background colors of the violin plots) are presented. From left to right, the average uncertainty estimate (spread) increases. As the spread increases, so does the estimate of the error, spanning multiple orders of magnitude on each axis. Critically, the distributions are mostly non-overlapping, meaning the uncertainty estimate can be used to produce a robust estimate for the actual error of the prediction. A more fine-grained presentation of the same data is presented in (d)-(f). These are error parity plots comparing the NNE spread with that of the actual error. To guide the eye, we show the best linear fit to the log-scaled data (solid), as well as four successive parallel lines offset by a half an order of magnitude (dashed). The majority ($>88$\%) of the points fall below the first of these upper bounds (half an order of magnitude above the best fit line), suggesting that most of the time, the error estimate given by this linear trend is an appropriate representation of the worst case scenario. We also note that even when using 10\% of the overall training set, the ability of the NNE to accurately quantify uncertainty is unaffected (see Fig.~\ref{fig:qm9-parity-p-resolved}).

It is also noteworthy to analyze why roughly 14\% of the data fall half an order of magnitude \textit{below} the best fit line. Even when making accurate predictions, each estimator will still predict slightly different values given the same input. These predictions can be accurate overall, but still produce noticeable values for an uncertainty estimate due to their slightly different predictions. This is a consequence of the way that neural networks train and make predictions. Each estimator finds some local minimum in its vast ``weight space", and each produces slightly different estimations even when the estimators and ensemble as a whole is making predictions to suitable accuracy. That said, under-confidence in a prediction is not nearly as problematic as over-confidence, and the amount of under-confidence as a function of $\hat{\sigma}^{(i)}_j$ decreases as uncertainty increases. In summary, our results suggest that the ensembles provide a robust, general upper bound for the error.

\subsection{Generalization test} \label{subsec:qm9-ml-generalized}

\begin{figure*}[!htb]
    \centering
    \includegraphics[width=2\columnwidth]{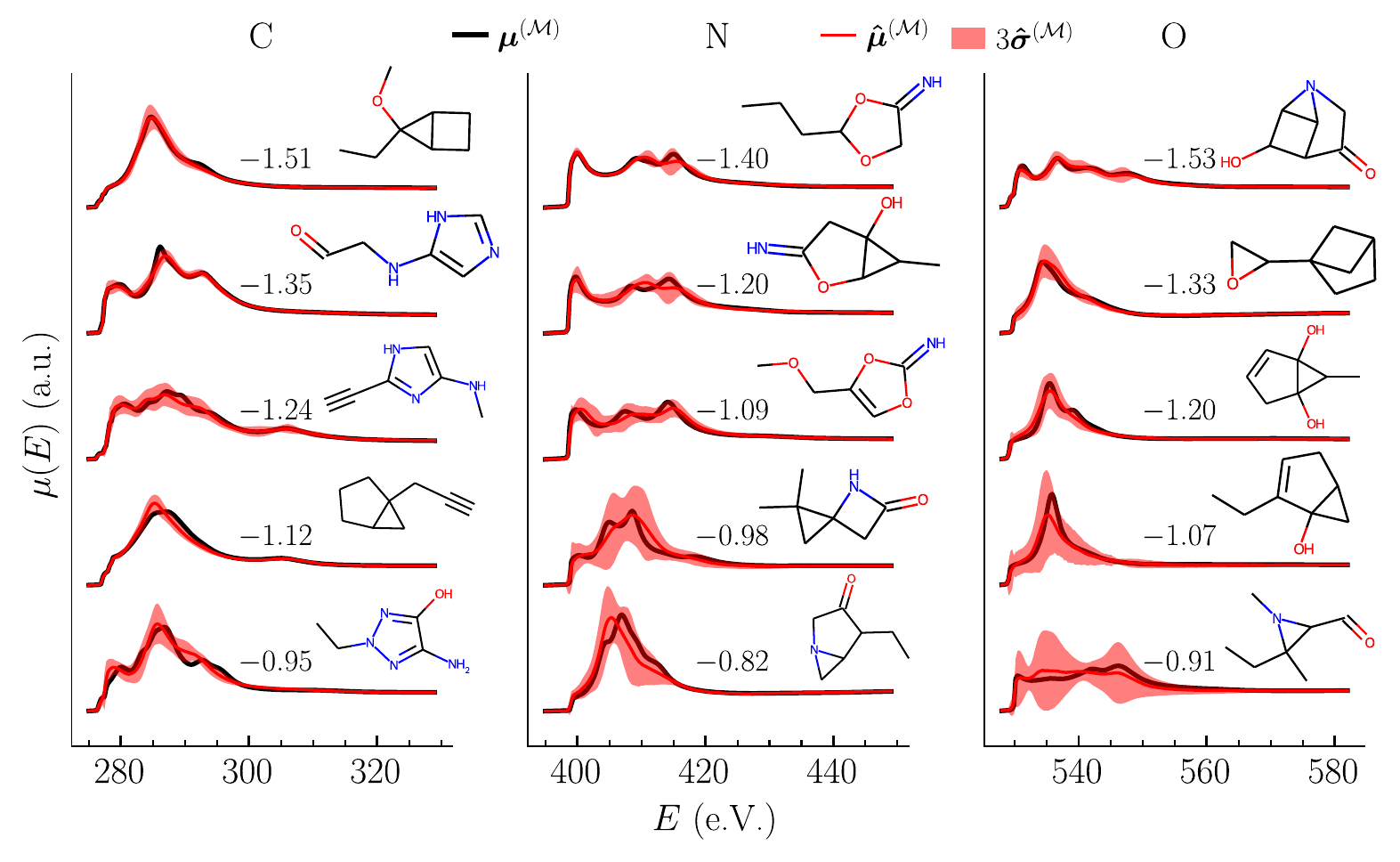}
    \caption{\label{fig:qm9-waterfall-CNO-examples} Waterfall plots similar to that of Fig.~\ref{fig:qm9_C_random_preds} showcasing examples from the middle of each pentile on the testing set predictions of $\mathcal{D}_A^\mathrm{G}$ for all $A=\{\mathrm{C}, \mathrm{N}, \mathrm{O}\}.$ The training sets used in these results consists of absorber-spectrum pairs originating from molecules with at most 6 heavy atoms, while the testing set is the entire subset of QM9 with 9 heavy atoms. The molecules of origin are shown in addition to the $\log_{10}$ error between the predicted molecular spectrum and the ground truth, along with the NNE predicted spread.}
\end{figure*}

The ability to generalize to previously unseen data is a key feature of any ML model. Generalization can be understood through the lens of data distributions: while the data used during testing must be unseen, the data used to train some model must, in a distributional sense, ``look like" the data it is expected to perform on. A simple way to test whether or not two sets of data are in-sample with respect to each other is to combine them and sample randomly. If a source of truth, e.g. a domain expert, can tell the distribution of origin given some random sample, then it is likely that the deployment case, which hopefully is represented by the testing set, is out-of-sample and will lead to poor performance. This is not a catch-all test (e.g. adversarial examples~\cite{goodfellow2014explaining}), but it is a useful thought experiment. For example, the testing sets as constructed randomly in Subsection~\ref{subsec:qm9-ml-random} were on balance, by the above definition, in-sample.

There is also a key difference between generalization and extrapolation which is worth noting. No ML models extrapolate beyond the information-theoretic union of the data and prior information they are trained on~\cite{mcPerspective}. Two examples of ``information-theoretic extrapolation" in our work would be (a) predicting on a molecule containing zwitterionic species and (b) predicting on un-relaxed structures. In case (a), the model has not seen any molecules with major charge gradients, and thus it will not understand how to treat those cases. Stated differently, it will neither understand which structural motifs correspond to a zwitterion nor how to treat them once detected. Similarly, in case (b), while it is possible that many structural configurations found in un-relaxed structures are captured in QM9 due to the large diversity of molecules in the dataset, there is no guarantee, since QM9 contains only relaxed geometries.

In this subsection, we push the boundaries of our NNEs to generalize to new data in a specific way, by training on sites from smaller molecules than what we test on. To do this, we train and cross-validate on subsets of the C, N and O databases in which there are at most 8 heavy atoms per molecule, and then test on sites originating from molecules containing 9 heavy atoms per molecule (see Subsection~\ref{subsec:qm9-data-prep} for details). The specifics of the training and evaluation procedures is identical to those presented in Subsection~\ref{subsec:qm9-ml-random}, except for the particular training/validation/testing split used.

Furthermore, the true test of the ability of the NNEs to generalize is to evaluate performance on \emph{molecular} spectra, defined as the average of the site-wise spectra [see Eq.~\eqref{eqn:average-mu}] (in Subsection~\ref{subsec:qm9-ml-random}, we only present results on site-spectra). For any $\mathcal{D}_A^\mathrm{G},$ a molecule is given by $\mathcal{M} \in \mathcal{D}_A^\mathrm{G},$ and is defined by a collection of sites. We define the subset of theses sites of atom type $A$ as $\mathcal{M}_A \subset \mathcal{M}.$ Given these definitions, the point-wise ground truth molecular XANES spectrum is,
\begin{equation} \label{eqn:average-mu}
    \mu_j^{(\mathcal{M}_A)} = \frac{1}{\abs{\mathcal{M}_A}} \sum_{i \in \mathcal{M}_A} \mu_j^{(i)},
\end{equation}
where $\abs{\mathcal{M}_A}$ is the number of absorbing sites of type $A$ in the molecule. Furthermore, the point-wise molecular XANES spectrum prediction is the average of each of the ensemble predictions for each site,
\begin{equation}
    \hat{\mu}_j^{(\mathcal{M}_A)} = \frac{1}{\abs{\mathcal{M}_A}} \sum_{i \in \mathcal{M}_A} \hat{\mu}_j^{(i)}.
\end{equation}
The estimate of the point-wise spread for the molecular XANES prediction can be calculated using propagation of errors,
\begin{equation}
    \hat\sigma_j^{(\mathcal{M}_A)} = \frac{1}{\abs{\mathcal{M}_A}} \sqrt{\sum_{i \in \mathcal{M}_A} \left[\hat\sigma_j^{(i)}\right]^2},
\end{equation}
with an analagous vector representation to that of Eq.~\eqref{eqn:boldsymbol-sigma-i},
\begin{equation} 
    \hat{\boldsymbol{\sigma}}^{(\mathcal{M}_A)} = \left[\hat\sigma_1^{(\mathcal{M}_A)}, \hat\sigma_2^{(\mathcal{M}_A)}, ..., \hat\sigma_M^{(\mathcal{M}_A)}\right].
\end{equation}
Finally, the error of the molecular spectrum is similarly given by a straightforward analog of Eqs.~\eqref{eqn:delta-i} and \eqref{eqn:delta-ij}. For brevity, we will often suppress the subscript $A$ where it is clear which atom type/database is being referred to.

We present waterfall plots of the ground truth molecular spectra, and the NNE predictions and spreads in Fig.~\ref{fig:qm9-waterfall-CNO-examples} in the $\mathcal{D}_A^\mathrm{G}$ databases. To demonstrate the ability of the NNE to generalize, we train only on absorbing site-spectrum pairs originating from molecules with at most 6 heavy atoms, but the presented testing set results come from absorbing site-spectrum pairs originating from molecules with 9 heavy atoms. This experiment demands an extreme degree of generalization from the NNE: the subsets of QM9 with only 6 heavy atoms is extremely small, containing only $\approx 10^3$ total structures, three orders of magnitude less than the testing set in this case (see Fig.~\ref{fig:qm9_N}). The dependence of the testing set error on the maximum number of atoms per molecule used in the training data (along with a similar analysis to that presented in Fig.~\ref{fig:qm9-parity}) is explored in Appendix~\ref{appendix:train-set-dependence}. The key result is that adding orders of magnitude more data results in only slight improvement in testing set error. For example, using the $\mathcal{D}_\mathrm{C}^\mathrm{G}$ dataset with up to 5 heavy atoms in the training set includes 437 data points, and produces an error of $\bar{\varepsilon} \approx 0.09.$ training with up to 8 heavy atoms uses a training set of 102253 data points, three orders of magnitude more data, and produces an error of $0.03$ on the same testing set. Using roughly $10^3$ times as much data produces only a factor of 3 improvement in the testing error. Combined with the results in Fig.~\ref{fig:qm9-waterfall-CNO-examples}, this indicates that the NNE is already able to generalize to larger molecules at a relatively low training cost, and is incredibly data-efficient. These results are further rigorously quantified in Appendix~\ref{appendix:train-set-dependence} (Fig.~\ref{fig:qm9-generalized-sigma-parity}).

In particular the results for $\mathcal{D}_\mathrm{C}^\mathrm{G}$ present with impressive accuracy given that each spectrum is an average over many carbon atoms (as many as 9 in one of the presented cases). Qualitatively, peak heights and locations are predicted to reasonable accuracy. In contrast, the results for $\mathcal{D}_\mathrm{N}^\mathrm{G}$ and $\mathcal{D}_\mathrm{O}^\mathrm{G}$ showcases where the NNE struggles to make accurate predictions for the number of absorbing atoms per molecule. This is largely due to there being far fewer training examples in these two cases relative to $\mathcal{D}_\mathrm{C}^\mathrm{G}.$ Even so, while some peaks are occasionally missed, the spectral trends are still reproduced, and uncertainties are captured. 

\subsection{Out-of-equilibrium geometry analysis} \label{subsec:distort}

\begin{figure}[!htb]
    \centering
    \includegraphics[width=\columnwidth]{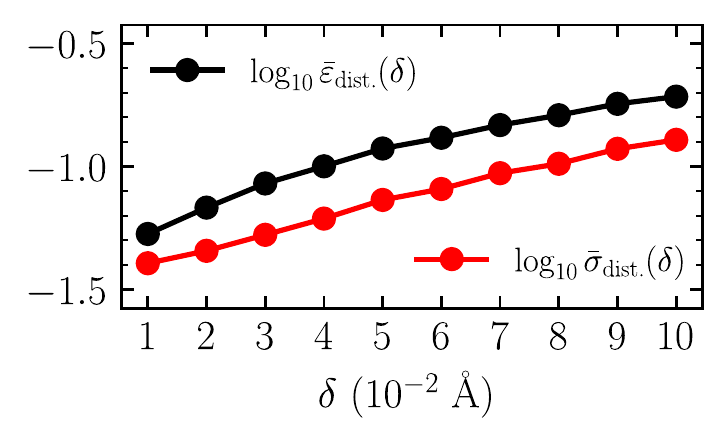}
    \caption{\label{fig:avg_distortion_results} The average $\log_{10}$ errors and uncertainties (across the 10 example molecules listed in Fig.~\ref{fig:qm9_C_random_preds} and 50 random distortions per molecule per value of delta) plotted as a function of $\delta.$}
\end{figure}

\begin{figure*}[!htb]
    \centering
    \includegraphics[width=2\columnwidth]{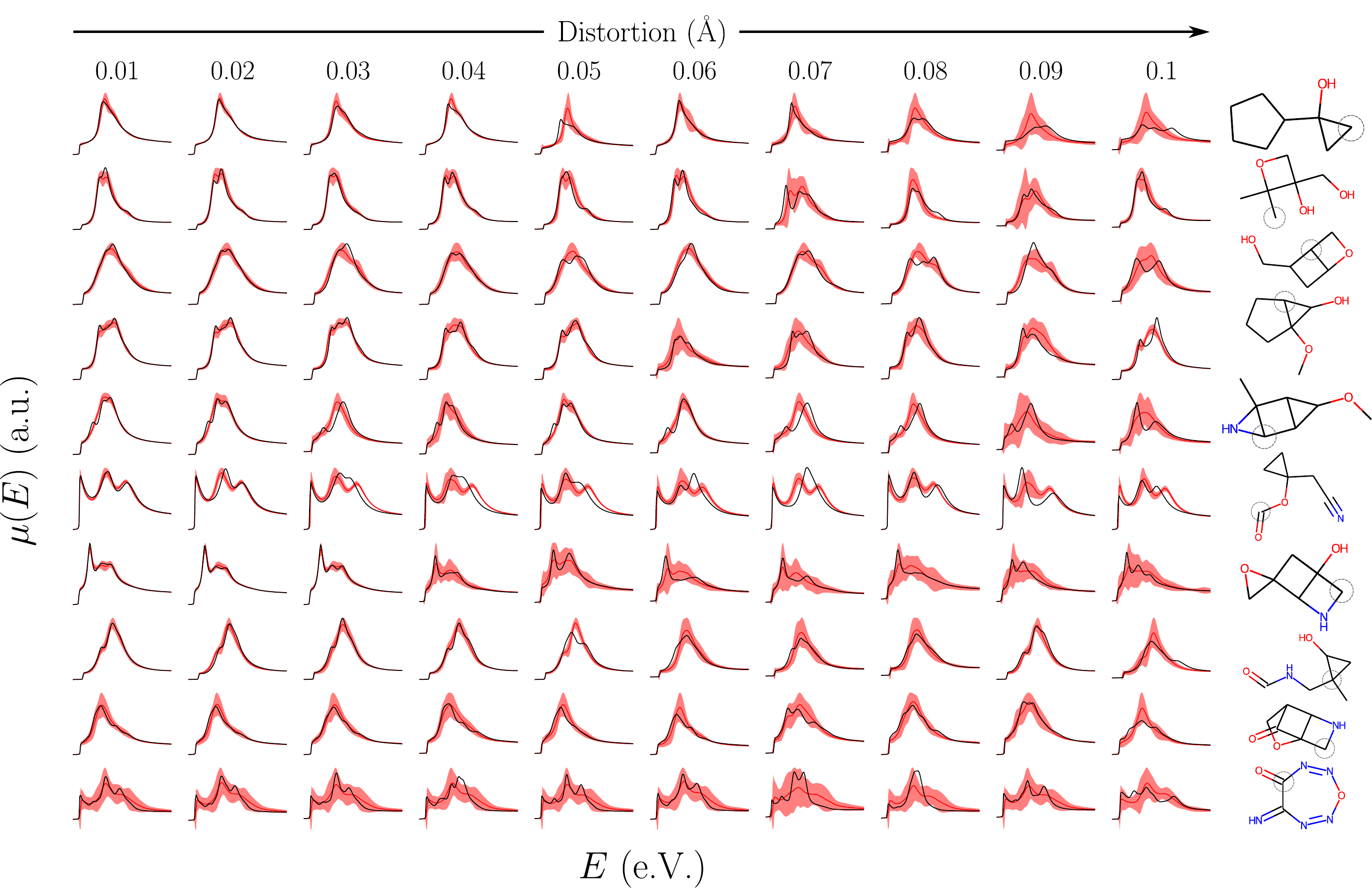}
    \caption{\label{fig:distortion_preds_waterfall} Predicted XANES spectra (red) and uncertainty estimations (red shaded regions) compared with the ground truth (black) for single examples for each site-$\delta$ pairing. Predictions on the same molecule-site pairs as Fig.~\ref{fig:qm9_C_random_preds} were computed in inference mode using models trained on the $\mathcal{D}^\mathrm{R}_\mathrm{C}$ dataset.}
\end{figure*}

To further study how the NNE responds to out-of-sample data, we used the 10 molecules corresponding to sites whose spectra were presented in Fig.~\ref{fig:qm9_C_random_preds}, and randomly distorted the geometries of the molecules to see how the NNE performs. Our procedure is as follows. First, given a distortion parameter $\delta,$ for each coordinate direction and atom in some molecule $\mathcal{M},$ a direction on the unit sphere is chosen at random, scaled by $\delta,$ and then used to perturb that atom's coordinate. For each of the 10 molecules and each value of $\delta,$ we sample 50 distorted molecules. Second, FEFF calculations are then run on each of these new geometries. Finally, the trained NNE used in the $\mathcal{D}^\mathrm{R}_\mathrm{C}$ experiments is then used to predict the XANES spectrum and spread for each of the site-molecule pairs.

We use $\delta \in \{0.01, 0.02, ..., 0.1\}$~\AA, and present averaged results analogous to Eqs.~\ref{eq:epsilon-bar} in Fig.~\ref{fig:avg_distortion_results} as a function of $\delta.$ Most importantly, as the average error increases across the average of all results for a given $\delta,$ so does the uncertainty measure. This trend is significant and covers roughly a half of an order of magnitude. This shows quantitatively that on average, the uncertainty estimate tracks how out-of-sample a dataset is with respect to its training data. However, it does not quantify the relative difference between a certain and uncertain prediction. To address this, we sample a single example for each molecule-$\delta$ pair, and plot the ground truth, NNE prediction and spreads in Fig.~\ref{fig:distortion_preds_waterfall}. Appendix~\ref{apdx:distortion} contains more fine-grained details relating to this analysis, including density parity plots of the errors and uncertainty measures (Fig.~\ref{fig:qm9_sigma_parity_distortion}), and a waterfall plot of distorted spectra (Fig.~\ref{fig:qm9_distortion_waterfall}).

While it is only a small sample of the dataset of distorted molecules, the results in Fig.~\ref{fig:distortion_preds_waterfall} clearly show that on average, the NNE is able to detect when distorted geometries are sufficiently out-of-sample to render a prediction inaccurate. Overall accuracy varies visually as the distortion increases, but the uncertainty estimate gets significantly larger after $\delta = 0.03.$ In most cases, this uncertainty stays relatively large compared to e.g. $\delta = 0.01,$ indicating the NNE recognizes that the geometry is likely unseen. We highlight that detecting when a geometry is out-of-sample is not a trivial task, since while the ACSF vectors are human-interpretable, they are not \textit{easily} so. Defining necessary heuristics to detect an out-of-sample geometry is likely not feasible. The fact that the NNE can perform this task verifies its potential usefulness in situations where detecting e.g. change points~\cite{flynn2019change} (where the statistical distribution of data changes) is required. This could be of particular use in active learning loops, where new training data is sampled based on the uncertainty measure of the ML model.

\section{Conclusions \& Outlook} \label{sec:conclusions}

In this work, we use NNEs to make quantitatively accurate predictions of molecular XANES spectra from local atomistic geometry, and to accurately quantify the uncertainty of those predictions under a variety of conditions. Simulating XANES spectra of a large number of molecules and clusters at a high level of theory is computationally demanding, and UA surrogate modeling provides an avenue for greatly accelerating the simulations while reliably quantifying model confidence. Often, for comparison with experimental measurements it requires an expensive averaging over a large number of structures, such as when computing a thermal average, which necessitates the use of surrogate model acceleration. We anticipate that the NNE approach, and UA modeling in general, can be particularly useful for large, complex systems, such as the dynamical evolution of protein structures, organic liquids and solvated molecules, allowing users to make predictions efficiently and with confidence.

Although our work falls into the space of ML-driven spectral function prediction, an accurate surrogate also has important implications for the inverse problem, where physical descriptors are extracted from the spectral function. For example, one strategy to solve the inverse problem in XAS is to identify candidate structures that produce results \textit{consistent} with a target. This is more broadly known as structure refinement. Various sampling methods are widely used for this purpose, such as Reverse Monte Carlo~\cite{mcgreevy1988reverse,mcgreevy2001reverse} and genetic algorithms~\cite{whitley1994genetic}. Combining these sampling methods with an accurate and efficient forward surrogate model opens new avenues to tackle the inverse problem.

Beyond the problems presented in this manuscript, we believe that the general principles of UQ methods could have broad implications not only for the case of datasets from \textit{in silico} experiments, but also for laboratory measurements in experimental science. While modeling aleatoric noise in experimental data is required for statistically robust predictions, UQ techniques can also be applied to, e.g., quality assurance and control. For example, the detection and elimination of data that results from a variety of experimental sources such as misalignment and errors in control settings (similar to Appendices~\ref{appendix:ensemble-prediction-details} and \ref{apdx:outliers}). UQ-enabled models could also be useful for predicting experiment-quality data. As long as noise and other sources of uncertainty can be \textit{accurately} modeled, it would mitigate the risk otherwise posed by the immense challenge of modeling experiments with these types of data-driven techniques.

As problems become more complicated, and calculations become more expensive (and thus the stakes become higher), ensuring model confidence becomes evermore important. Conveniently, there exists a vast array of UA models and methodologies for quantifying uncertainty, each with their own strenghts and weaknesses. In the case of NNEs, the cost of training multiple models is more than worth the payoff. The application of UQ techniques to vector targets, and to a wider variety of problems in the physical sciences is certainly still an open problem in general. However, we have found that one can gain a significant amount of utility through the straightforward use of a NNE: an ensemble of independent estimators. In this case, if you can train one model, you can train a sufficient number of models to produce a reasonable measure of uncertainty with a low overhead.

In conclusion, our case study demonstrates the robust performance of a NNE with uncertainty quantification for predicting complex targets (XANES spectra of small molecules) from the descriptors of local chemical and structural information. More generally, this expands the scope of uncertainty aware machine learning methods to the case of predicting vector quantities in physical modeling, an area that is largely unexplored to date. UQ modeling offers compelling advantages over traditional more boilerplate machine learning techniques at an acceptable cost.

\section*{Data Availability}

All software used in this work can be found open-source at \href{https://github.com/AI-multimodal/XAS-NNE}{github.com/AI-multimodal/XAS-NNE}. All data used in this work, including FEFF spectra input/output files, featurized data, and the neural network ensembles can be found open access at \href{https://doi.org/10.5281/zenodo.7554888}{\code{doi.org/10.5281/zenodo.7554888}}.

\begin{acknowledgements}
MRC would like to thank Nongnuch Artrith and Alexander Urban for helpful discussions regarding active learning. This research is based upon work supported by the U.S. Department of Energy, Office of Science, Office Basic Energy Sciences, under Award Number FWP PS-030. This research also used theory and computational resources of the Center for Functional Nanomaterials, which is a U.S. Department of Energy Office of Science User Facility, and the Scientific Data and Computing Center, a component of the Computational Science Initiative, at Brookhaven National Laboratory under Contract No. DE-SC0012704. This project was supported in part by the U.S. Department of Energy, Office of Science, Office of Workforce Development for Teachers and Scientists (WDTS) under the Science Undergraduate Laboratory Internships Program (SULI).
\end{acknowledgements}

\begin{widetext}

\appendix

\section{Database construction}

\subsection{ACSF feature vector details} \label{appendix:acsf}

Molecular geometry files were read directly from the QM9~\cite{ramakrishnan2014quantum} database, which is freely available for download at \href{http://quantum-machine.org/datasets/}{quantum-machine.org/datasets/}. We utilize the Pymatgen~\cite{ong2013python}, DScribe~\cite{dscribe} and ASE~\cite{larsen2017atomic} libraries to construct our ACSF feature vectors. Zwitterionic molecules, those which contain an equal number of positively and negatively charged motifs, are discarded and not included in the machine learning databases. The initialization of the \code{ACSF} object is shown below.

\begin{lstlisting}[linewidth=\columnwidth,frame=none]
from dscribe.descriptors import ACSF  # v 1.2.1
neighbors = ["H", "C", "O", "N", "F"]
rcut = 6.0  # Angstroms
g2_params = [
    [1.0, 0],
    [0.1, 0],
    [0.01, 0]
]
g4_params=[
    [0.001, 1.0, -1.0],
    [0.001, 2.0, -1.0],
    [0.001, 4.0, -1.0],
    [0.01, 1.0, -1.0],
    [0.01, 2.0, -1.0],
    [0.01, 4.0, -1.0],
    [0.1, 1.0, -1.0],
    [0.1, 2.0, -1.0],
    [0.1, 3.0, -1.0]
]
acsf = ACSF(
    species=neighbors,
    rcut=rcut,
    g2_params=g2_params,
    g4_params=g4_params
)
\end{lstlisting}

We iterate through every possible atom in each molecule, and construct the atom's ACSF vector if it matches an absorbing atom used in this work (C, N, or O).

\subsection{Spectra target vector details} \label{appendix:spectra}

Site-wise spectra for each of C, N and O absorbing atoms were computed using the FEFF9 code~\cite{rehr2010parameter}. A common preamble to a FEFF calculation is shown below. Particularly, we use a corehole approximation at the Random Phase Approximation (RPA)-level of theory, full multiple scattering up to 9~\AA, and self-consistency up to 7~\AA.

\begin{lstlisting}[linewidth=\columnwidth,frame=none]
TITLE ...

EDGE      K
S02       1.0
COREHOLE  RPA
CONTROL   1 1 1 1 1 1

XANES     4 0.04 0.1

FMS       9.0
EXCHANGE  0 0.0 0.0 2
SCF       7.0 1 100 0.2 3
RPATH     -1
\end{lstlisting}

Once the initial spectral databases were constructed, we screen for extreme outliers or unphysical spectra using methods similar to those described in Ref.~\onlinecite{carbone2019classification} (though we note that these screening procedures are not entirely robust, see the discussion corresponding to Fig.~\ref{fig:qm9_O_fail}). Spectra are then interpolated onto common grids using cubic splines. The grids for each absorbing atom type are shown below using Python+NumPy.

\begin{lstlisting}[linewidth=\columnwidth,frame=none]
import numpy as np
M = 200
grids = {
    "C": np.linspace(275, 329, M),
    "N": np.linspace(395, 449, M),
    "O": np.linspace(528, 582, M)
}
\end{lstlisting}

\section{Training details} \label{appendix:training-details}
In this appendix, we highlight the important details of our training procedures which can be used to reproduce the work presented in this manuscript. All training was performed on Tesla V100 GPUs using Pytorch+PyTorch Lightning, and the summary of the training and software used for our machine learning pipeline are shown in Table~\ref{tab:apdx:training-details}.

\begin{table}[!htb]
\caption{\label{tab:apdx:training-details}
Machine software and hardware details.
}
\begin{ruledtabular}
\begin{tabular}{ll}
GPU & Tesla V100-SXM2-32GB \\
CUDA & 11.4 \\
PyTorch~\cite{pytorch} & 1.11.0+cu113 \\
PyTorch Lightning~\cite{Falcon_PyTorch_Lightning_2019} & 1.6.4
\end{tabular}
\end{ruledtabular}
\end{table}

Each estimator, a FFNN, was trained independently. Each estimator in the ensemble was randomly initialized, with a minimum of 4 layers, a maximum of 20 layers, a minimum of 160 neurons/layer, and a maximum of 200 neurons/layers. The Adam optimizer, L1 loss function, Leaky ReLU activations (except the last layer, which is a softplus) and batch normalization (except the last layer) were used for every instance of training.

During training, learning rates were multiplied by a factor of $0.95$ after 20 successive epochs in which the validation loss failed to decrease. A maximum of 2000 epochs proved sufficient with early stopping criteria monitoring when the validation loss plateaued during training (with a patience of 100 epochs).

\section{Ensemble prediction details} \label{appendix:ensemble-prediction-details}

In Subsection~\ref{subsec:qm9-ml-random}, we defined the set of estimator indexes to be $\mathcal{E},$ and the set of estimator indexes corresponding to ``reasonable" predictions to be $\mathcal{E}(i).$ We now define precisely how we determine whether or not a prediction is reasonable.

We utilize three criteria to screen for unreasonable predictions:
\begin{enumerate}
    \item For a set of predicted spectra (with fixed $i$) $\{\hat{\boldsymbol{\mu}}^{(i, k)}\}_{k=1}^{\abs{\mathcal{E}}},$ we define the estimator spread $\hat{\boldsymbol{\sigma}}^{(i)}.$ Any spectra in which 70\% of the total grid points fall outside a region defined by $\hat{\boldsymbol{\mu}}^{(i)} \pm 2\hat{\boldsymbol{\sigma}}^{(i)}$ are discarded.
    \item Given a predicted spectrum $\hat{\boldsymbol{\mu}}^{(i, k)},$ if any point $\hat{\mu}^{(i, k)}_j$ is greater than 20 (a.u.; roughly an order of magnitude greater than the largest spectral intensity in our datasets) that prediction is discarded.
    \item If more than half of the predictions $\hat{\mu}^{(i, k)}_j$ for a given $i, k$ fall below an intensity of 0.05, that prediction is discarded.
\end{enumerate}

These three rules are purely empirical and capture three different kinds of model failure scenarios, each of which is completely independent of the ground truth data. In point 1, we screen for statistical outliers. The spread of the estimator predictions is treated as a Gaussian standard deviation (we note again that this treatment is purely empirical), and any spectra in which a substantial portion of the predicted values fall outside of a $\approx 95\%$ confidence interval are discarded. Points 2 and 3 screen on physical grounds. We know from experience and quantum mechanical principles that the FEFF code will only output predictions of certain intensities. Model failure situations routinely included single estimators predicting intensities of $\mu > 100,$ which are clearly unphysical. Similarly, it is known that the XAS is mostly positive. While model outputs are hard-constrained to be non-negative by the Softplus activation functions, they can be approximately zero. If a sufficient portion of the spectrum is close to zero, we know that prediction is unphysical and hence it is discarded.

The set of estimator indexes that correspond to reasonable predictions is thus defined as $\mathcal{E}(i),$ and is inherently dependent on the training example. It is worth highlighting that if estimator $k$ is discarded for training example $i,$ it likely will not be discarded for example $i' \neq i.$

\section{Training set size-dependence of $\mathcal{D}_A^\mathrm{R}$ and $\mathcal{D}_A^\mathrm{G}$} \label{appendix:train-set-dependence}

\begin{figure}[!htb]
    \centering
    \includegraphics[width=\columnwidth]{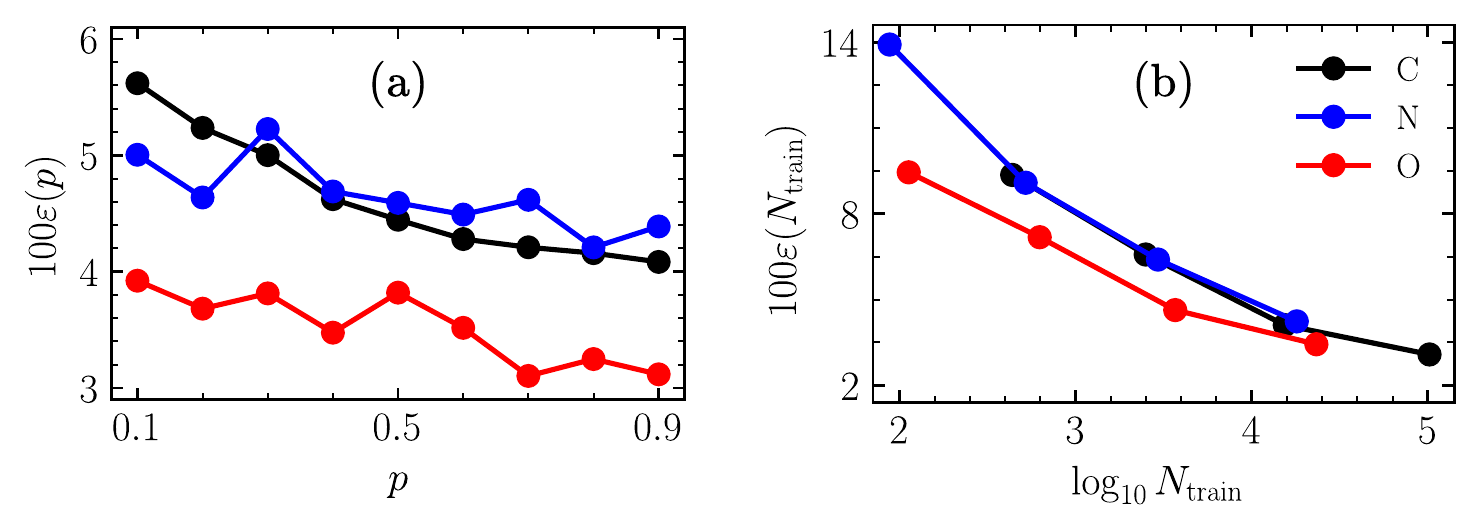}
    \caption{\label{fig:qm9-errors-as-p} (a) The testing set error on $\mathcal{D}_A^\mathrm{R}$ as a function of the downsampling proportion $p,$ which determines the random proportion of the training set actually used during training. (b) The testing set error on $\mathcal{D}_A^\mathrm{G}$ as a function of the training set size, $\abs{\mathcal{M}}$ (which is tied to the number of atoms per molecule in the training set).}
\end{figure}

\begin{figure*}[!htb]
    \centering
    \includegraphics[width=\columnwidth]{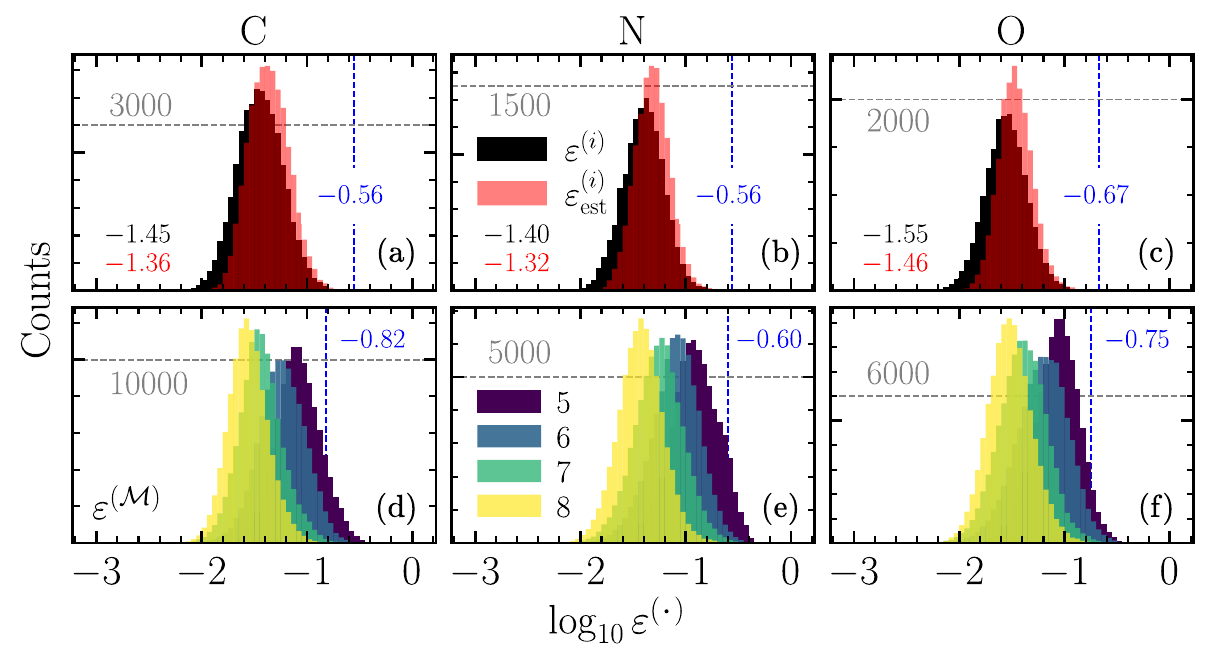}
    \caption{\label{fig:qm9-all-hists} Histograms of the $\log_{10}$ testing set error of the molecular predictions between the predicted and ground truth spectra for the $\mathcal{D}_A^\mathrm{R}$ and $\mathcal{D}_A^\mathrm{G},$ (a)-(c) and (d)-(f), respectively. Results in (a)-(c) showcase the distribution of ensemble errors (black) and distribution of the average estimator errors (red), which correspond to Eqs.~\eqref{eqn:delta-i} and \eqref{eqn:delta-i-est}, respectively. The median values for these results are also shown. Results in (d)-(f) are resolved by the number of heavy atoms/molecule $\abs{\mathcal{M}} \in \{5, 6, 7, 8\}$ in the training set. All plots show the respective ``dummy model" prediction (average of the testing set spectra) in dashed blue.}
\end{figure*}

\begin{figure*}[!htb]
    \centering
    \includegraphics[width=\columnwidth]{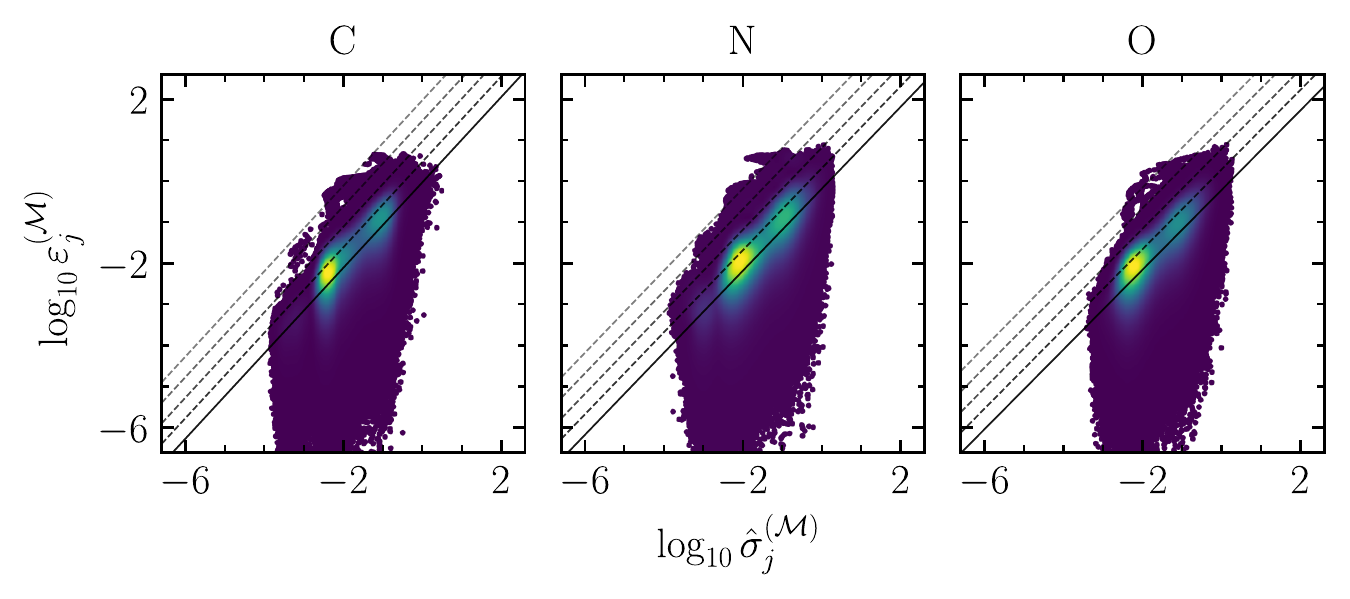}
    \caption{\label{fig:qm9-generalized-sigma-parity} Parity density plot for the $\mathcal{D}_A^\mathrm{G}$ datasets in which molecules with at most 6 heavy atoms were used for training.}
\end{figure*}

\begin{figure}[!htb]
    \centering
    \includegraphics[width=0.6\columnwidth]{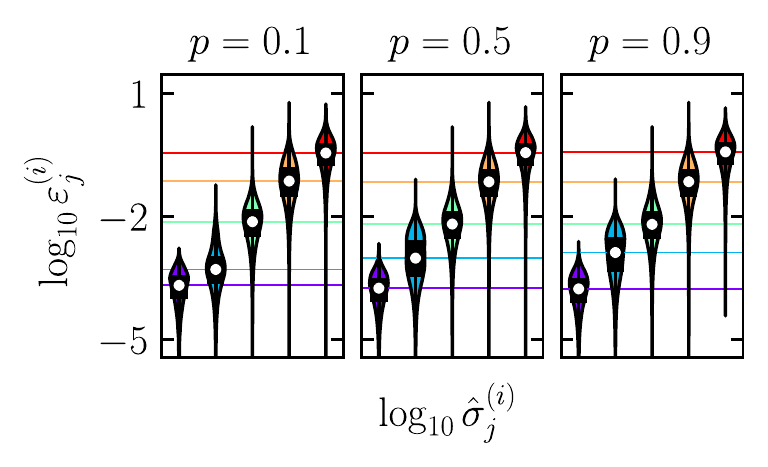}
    \caption{\label{fig:qm9-parity-p-resolved} Violin plots similar to that of Fig.~\ref{fig:qm9-parity} in the main text (with the same bins), but resolved by the proportion of the training set size, $p.$ Results are presented for $\mathcal{D}_\mathrm{C}^\mathrm{R}.$}
\end{figure}

A useful sanity check when performing any ML modeling is to ensure that error decreases as the training set size increases. For any single estimator, this is usually tested by downsampling the training set by some proportion $p$ and evaluating on a fixed testing set. The value for $p$ is then scanned and the testing set error as a function of $p$ evaluated. We evaluate this metric for both $\mathcal{D}_A^\mathrm{R}$ and $\mathcal{D}_A^\mathrm{G},$ by randomly downsampling and showcasing how testing set error changes as a function of the number of atoms per molecule used during training. 

Instead of training a single estimator, we evaluate these metrics on the ensembles. For the random partitioning datasets $\mathcal{D}_A^\mathrm{R},$ we randomly downsample the fixed training set by proportion $p$ independently for each of the 30 estimators, train them on the $pN_\mathrm{train}$ training examples, and evaluate on the fixed testing set (the same testing set used in the random partitioning dataset, see Subsection~\ref{subsec:qm9-ml-random}). Each ensemble was randomly initialized for each $p.$ Ensemble predictions are made in accordance with the protocol described in Appendix~\ref{appendix:ensemble-prediction-details}, and the results averaged over all $N_\mathrm{test}$ testing examples. Similarly, for the generalization test databases $\mathcal{D}_A^\mathrm{G},$ we follow the procedure as outlined in Subsection~\ref{subsec:qm9-ml-generalized}, and for each instance of the number of atoms per molecule, compute the amount of training data used. All of these results are plotted in Fig.~\ref{fig:qm9-errors-as-p}.

Immediately, it is clear that the testing set error \textit{trend} is decreasing with increasing $p$ and with increasing amounts of training data, exactly as expected. However in Fig.~\ref{fig:qm9-errors-as-p}(a), the decrease is not monotonic for the N and O datasets, though it is for the C dataset. Given that the C dataset is much larger (see Fig.~\ref{fig:qm9_N}), it is more probable that even small samples of the training dataset capture general trends, even for small $p.$ The N and O datasets are significantly smaller, and hence for small $p$ it is less likely that significant trends are captured on balance. Finally, the relative decrease in the testing error as a function of $p$ is surprisingly small. For example, the percent change in the N results from $p=0.1$ to $0.9$ is only roughly 12\%. While this is a relatively small change (see e.g. the Supplementary Material in Ref.~\onlinecite{carbone2020machine}), it is actually an encouraging result. It is likely that failures of individual estimators in certain regions of the input parameter space at small $p$ are compensated for by other estimators in the ensemble, making the overall inference procedure at low $p$ surprisingly robust.

In Fig.~\ref{fig:qm9-errors-as-p}(b), the trend is much more clear because the amount of training data spans multiple orders of magnitude. However, a key takeaway is that the decrease in the error is incredibly small relative to the amount of training data used: roughly a single order of magnitude compared to 3 orders of magnitude increase in the amount of training data. As discussed in the main text, this is a testament to the NNE method's ability to generalize when trained on absorbing site data, and the diversity of the chemical space of QM9's molecules containing even $\sim 10^3$ heavy atoms per molecule.

We present the error histograms for most of the relevant training procedures in this work in Fig.~\ref{fig:qm9-all-hists}. Subfigures (a)-(c) depict the $\log_{10}$ site-wise testing set error distributions for $\mathcal{D}_A^\mathrm{R},$ both for the ensemble-averaged error (black) and the average single estimator error (red). One can see clearly that the ensemble predictions are systematically better than any single estimator on average, as expected of ensemble-based models. Subfigures (d)-(f) showcase the $\log_{10}$ molecular testing set error distributions for $\mathcal{D}_A^\mathrm{G},$ as a function of the total number of atoms per molecule used during training (see legend). Figure~\ref{fig:qm9-generalized-sigma-parity} is the analog of Fig.~\ref{fig:qm9-parity} for the molecular data. It shows the same trends and overall behavior as the aforementioned site-spectra figure in the main text.

Finally, we present a qualitative analysis of how the error and standard deviations correlate as a function of training set size $pN_\mathrm{train}$ in Fig.~\ref{fig:qm9-parity-p-resolved}. Even for $p=0.1,$ we see that the correlation between the average errors and standard deviations remains intact. This is an indication that even with limited training data, uncertainty-quantifying models can still accurately gauge when they are out of sample, and provide a reasonable estimate as to their own uncertainty.

\section{Explanation of Fig.~\ref{fig:qm9-parity} outliers} \label{apdx:outliers}

\begin{figure}[!htb]
    \centering
    \includegraphics[width=0.5\columnwidth]{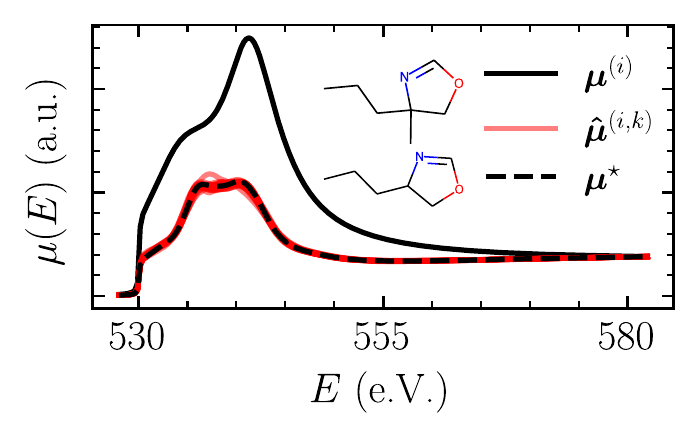}
    \caption{\label{fig:qm9_O_fail} An example from $\mathcal{D}_\mathrm{O}^\mathrm{R}$ of the NNE making an accurate prediction even when the source of ground truth used to train the ensemble was incorrect. The supposed ground truth (solid black) and the predictions of each estimator (red) are compared to the closest spectrum to $\boldsymbol{\hat{\mu}}^{(i)}$ in the training database (dashed black, $\boldsymbol{\mu}^\star$). The $\log_{10}$ error between $\boldsymbol{\mu}^\star$ and the prediction $\boldsymbol{\hat{\mu}}^{(i)}$ is $-2.16,$ corroborating the relatively low $\log_{10}$ uncertainty estimate $-1.79.$}
\end{figure}

The C and O results in Fig.~\ref{fig:qm9-parity} present with some interesting outlier behavior (for example, the data above the 4th dashed lines). These patterns clearly indicate something has gone systematically wrong, and as such, as a pedagogical exercise we investigate the cause of such significant underestimation of the errors. As an example, we look at the O results; it turns out that most of the outliers (captured by taking $\log_{10} \hat{\sigma}_j^{(i)} < -2$ and $\log_{10} \hat{\varepsilon}_j^{(i)} > -1$ in Fig.~\ref{fig:qm9-parity}) come from a single spectrum. This failure case is plotted in Fig.~\ref{fig:qm9_O_fail}.

The significant underestimation of the error is visually obvious by comparing the spread of the predictions (red) with the supposed ground truth (black). However, this ground truth spectrum is actually an outlier with respect to other training data, suggesting the possibility that the ground truth is incorrect. To test this hypothesis, we also plot the closest ground truth spectrum to the mean of the predicted spectra from the \textit{training} set (dashed black, $\boldsymbol{\mu}^\star$). These two spectra are in almost perfect agreement. We then compare the molecular structures of the inputs, which are also shown in Fig.~\ref{fig:qm9_O_fail}. The SMILES string corresponding to the ground truth and best training set example, \code{CCCC1(C)COC=N1} and \code{CCCC1COC=N1}, respectively, are chemically almost identical, differing by a single methyl group. Additionally, the top 3 closest spectra to the ensemble-averaged prediction all contain a \code{N=C-O} motif contained in a 5-membered ring, strongly suggesting that the FEFF calculation used to generate the ground truth spectrum failed to properly converge, and indicating that the NNE prediction, and low uncertainty estimate, are actually correct.

In order to confirm this hypothesis, we performed two sanity checks. First, we re-ran the FEFF calculations to see if the convergence failure was systemic. Second, we double checked that a different computational spectroscopy software (we chose to use the Vienna Ab-inito Simulation Package, or VASP~\cite{kresse1999ultrasoft}), produced similar spectra for these absorbing sites as well. In summary, our original FEFF calculation failed to converge, the re-run produced the expected result (similar to $\boldsymbol{\mu}^\star$ in Fig.~\ref{fig:qm9_O_fail}), and the two VASP calculations (VASP calculations were performed using PBE pseudopotentials a $2\times 2 \times 2$ grid for the k-points, and a square supercell volume of $20 \times 20 \times 20$; all quantities are in units of \AA) also produced qualitatively similar spectra, confirming that indeed the O-XANES spectrum of each of these molecules is essentially the same, and that the NNE prediction is correct.

Of course, for a real world deployment scenario, any piece of ground truth data that ends up being unphysical would be removed from the training datasets. However, in this case, we highlight that the NNE model was robust to these outliers. Future work could be dedicated to exploring how robust NNEs or related methods are for outlier detection in a database in general, especially in cases where it is suspected the source of truth can actually be incorrect.

\section{Supplementary analysis of out-of-equilibrium distortion tests} \label{apdx:distortion}

\begin{figure}[!htb]
    \centering
    \includegraphics[width=1.0\columnwidth]{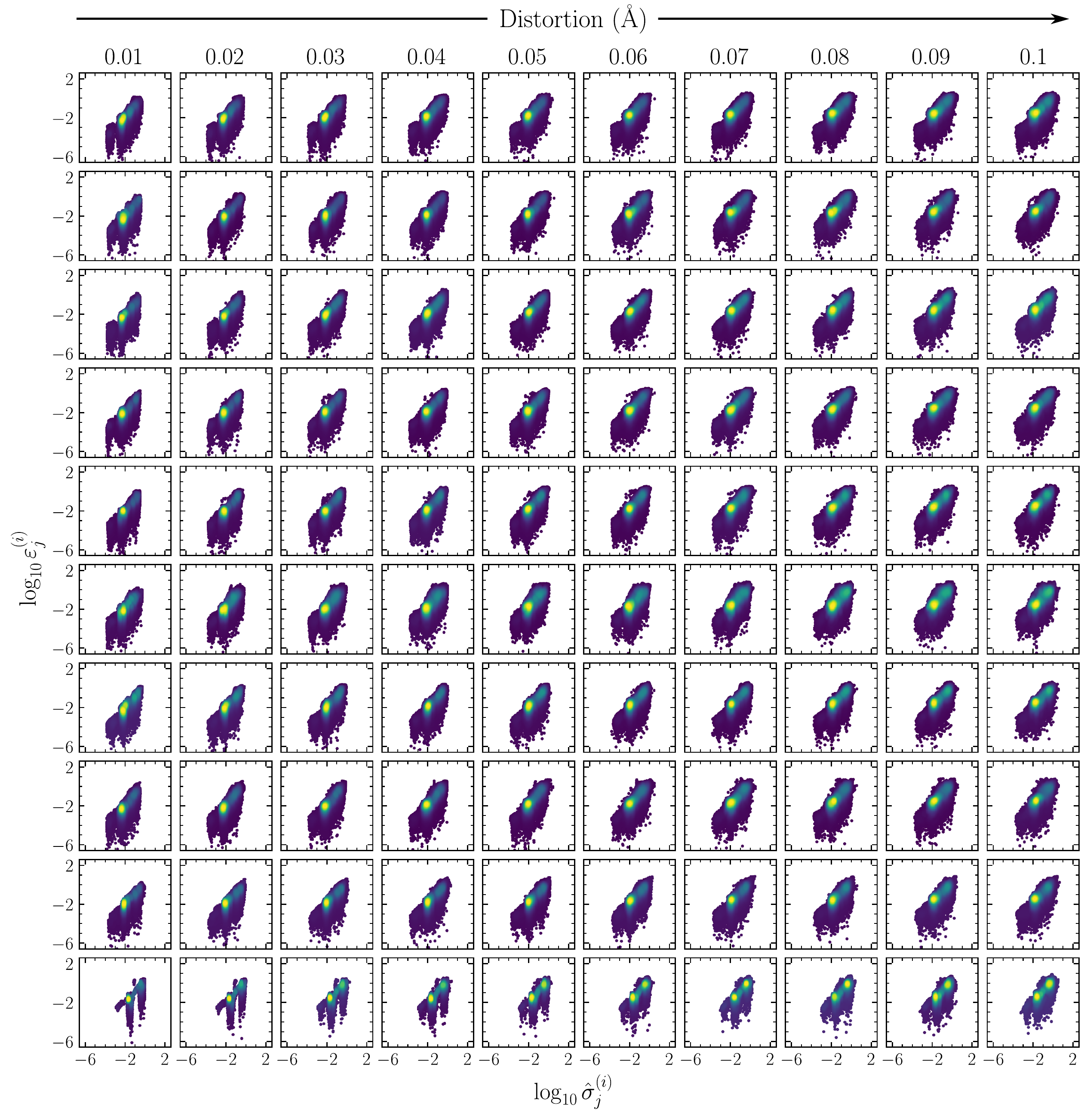}
    \caption{\label{fig:qm9_sigma_parity_distortion} Parity density plots similar to those in Figs.~\ref{fig:qm9-parity} and \ref{fig:qm9-generalized-sigma-parity}. Each column is indexed by a distortion value, $\delta,$ which is given in units of Angstroms and represents the maximum value of uniformally sampled random noise used to distort the locations of each atom along the $x,$ $y$ and $z$ axes. Approximately 50 random samples for each value of $\delta$ were used (less than 1\% of the time a FEFF calculation would fail). Each row corresponds to a different site from the $\mathcal{D}_\mathrm{C}^\mathrm{R}$ dataset, specifically those of Fig.~\ref{fig:qm9_C_random_preds}.}
\end{figure}

\begin{figure}[!htb]
    \centering
    \includegraphics[width=1.0\columnwidth]{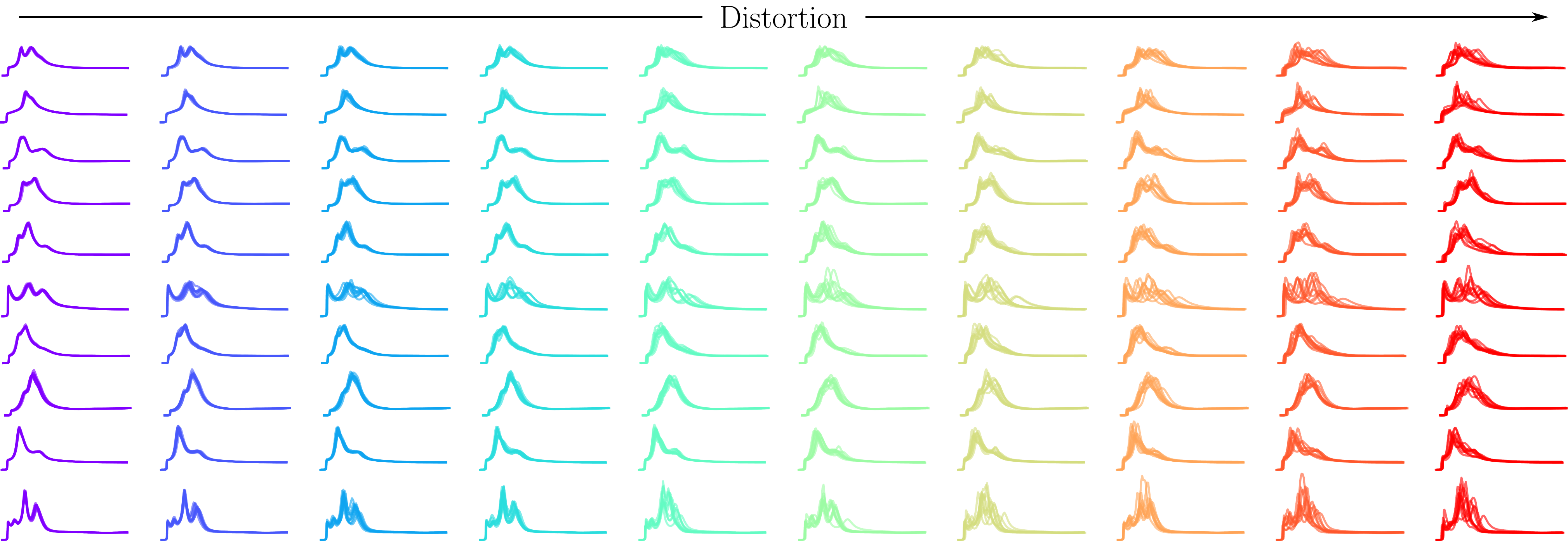}
    \caption{\label{fig:qm9_distortion_waterfall} Waterfall plot of the XANES spectra corresponding to the distorted geometries as outlined in Subsection~\ref{subsec:distort}. Each row corresponds to a specific site on a different molecule (the molecules are in the same order as presented in Fig.~\ref{fig:qm9_C_random_preds}), and each column corresponds to a different degree of distortion, $\delta \in \{0.01, 0.02, ..., 0.1\}$~\AA. }
\end{figure}

In this Appendix, we present two useful figures for the analysis of the out-of-equilibrium geometry tests. First, in Fig.~\ref{fig:qm9_sigma_parity_distortion}, we present a parity plot of the point-wise $\log_{10}$ errors vs. the $\log_{10}$ uncertainty estimates. The same positive linear trend found in Figs.~\ref{fig:qm9-parity} and \ref{fig:qm9-generalized-sigma-parity} is realized here. Additionally, it is subtle, but as the distortion amount increases, so does the overall error \textit{and} uncertainty estimate, continuing to substantiate the results in Subsection~\ref{subsec:distort}.

Second, to provide an idea of what distorted spectra look like, in Fig.~\ref{fig:qm9_distortion_waterfall}, we showcase a waterfall plot of random sampling of the database of distorted spectra, resolved once again by values of $\delta$ and molecule (in the same order as the previous related figures). As a result of the geometry distortion, every molecule begins to exhibit significant new spectral trends. Given that these new geometries are necessarily not in their equilibrium state, they will be much more challenging (if not infeasible) for the NNE to predict, and provide a good test for the NNE's ability to quantify epistemic uncertainty.

\end{widetext}

\end{document}